%% file: main.tex
\documentclass[screen, acmsmall]{acmart}

\AtBeginDocument{%
  \providecommand\BibTeX{{%
    Bib\TeX}}}

\setcopyright{acmlicensed}
\copyrightyear{2018}
\acmYear{2018}
\acmDOI{XXXXXXX.XXXXXXX}

\acmConference[Conference acronym 'XX]{Make sure to enter the correct
  conference title from your rights confirmation emai}{June 03--05,
  2018}{Woodstock, NY}
\acmISBN{978-1-4503-XXXX-X/18/06}

\usepackage{amsmath,amsfonts}%
\usepackage{graphicx}
\usepackage{textcomp}
\usepackage{xcolor}
\usepackage{booktabs}
\usepackage{listings}
\usepackage{multirow}
\usepackage{caption}
\usepackage{subcaption}
\usepackage{framed}

\usepackage{soul}
\usepackage[utf8]{inputenc}
\usepackage[T1]{fontenc}
\usepackage{microtype}
\usepackage{rotating}
\usepackage{paralist}
\usepackage{tabularx}
\usepackage{multicol}
\usepackage{pbox}
\usepackage{makecell}
\usepackage{enumitem}	
\usepackage{colortbl}
\usepackage{hyperref}
\usepackage{pifont}
\usepackage{xspace}
\usepackage{url}
\usepackage{tikz}
\usepackage{fontawesome}
\usepackage{lscape}
\usepackage{color}
\usepackage{xcolor}
\usepackage{anyfontsize}
\usepackage{comment}
\usepackage{soul}
\usepackage{gensymb}
\usepackage{multibib}
\usepackage[most]{tcolorbox}
\usepackage{balance}
\usepackage{footmisc}

\usepackage{listings, soul}
\usepackage[ruled,linesnumbered]{algorithm2e}
\def\BibTeX{{\rm B\kern-.05em{\sc i\kern-.025em b}\kern-.08em
    T\kern-.1667em\lower.7ex\hbox{E}\kern-.125emX}}

\usepackage{xspace}

\newcommand{\dc}{\textsc{DeepCrime}\xspace} 
\newcommand{\AutoT}{\textsc{AutoTrainer}\@\xspace}
\newcommand{\dfd}{\textsc{DeepFD}\xspace}

\definecolor{darkblue}{rgb}{0.0,0.0,0.6}

\newboolean{showcomments}
\setboolean{showcomments}{true}         
\ifthenelse{\boolean{showcomments}}
  {\newcommand{\nb}[2]{
  \fbox{\bfseries\sffamily\scriptsize#1}
     {\sf\small$\blacktriangleright$\textit{\textcolor{red}{#2}}$\blacktriangleleft$}
   }
  }
  {\newcommand{\nb}[2]{}
   
  }

\newcommand{\DL}{\textsc{DeepLocalize}\xspace}
\newcommand{\DD}{\textsc{DeepDiagnosis}\xspace}
\newcommand{\UM}{\textsc{UMLAUT}\xspace}
\newcommand{\AT}{\textsc{AutoTrainer}\xspace}
\newcommand{\NL}{\textsc{Neuralint}\xspace}

\newcommand{\etal}{\textit{et al.}\xspace}

\newcommand{\COMMENT}[1]{}

\definecolor{box-white}{cmyk}{0, 0, 0, 0, 0}
\definecolor{codegreen}{RGB}{0, 160, 0}
\definecolor{codered}{RGB}{160, 0, 0}
\definecolor{boxcolorblue}{RGB}{166,206,227}
\definecolor{boxcolorred}{RGB}{251,154,153}
\definecolor{boxcolorpurple}{RGB}{244,164,251}
\definecolor{boxcolorgreen}{RGB}{179,226,205}

\begin{document}

\title{Fault Localisation and Repair for DL Systems: An Empirical Study with LLMs}

\author{Jinhan Kim}
\orcid{0000-0002-0140-7908}
\affiliation{
  \institution{Università della Svizzera italiana (USI)}
  \city{Lugano}
  \country{Switzerland}}
\email{jinhan.kim@usi.ch}

\author{Nargiz Humbatova}
\orcid{0000-0002-3037-8368}
\affiliation{
  \institution{Università della Svizzera italiana (USI)}
  \city{Lugano}
  \country{Switzerland}
}
\email{nargiz.humbatova@usi.ch}

\author{Gunel Jahangirova}
\orcid{0000-0002-1423-1083}
\affiliation{
 \institution{King's College London}
  \city{London}
 \country{UK}
}
 \email{gunel.jahangirova@kcl.ac.uk}

\author{Shin Yoo}
\orcid{0000-0002-0836-6993}
\affiliation{
  \institution{KAIST}
  \city{Daejeon}
  \country{Republic of Korea}}
\email{shin.yoo@kaist.ac.kr}

\author{Paolo Tonella}
\orcid{0000-0003-3088-0339}
\affiliation{
  \institution{Università della Svizzera italiana (USI)}
  \city{Lugano}
  \country{Switzerland}
  }
\email{paolo.tonella@usi.ch}

\renewcommand{\shortauthors}{Kim et al.}

\begin{abstract}

  Numerous Fault Localisation (FL) and repair techniques have been proposed to address faults in Deep Learning (DL) models. However, their effectiveness in practical applications remains uncertain due to the reliance on pre-defined rules. This paper presents a comprehensive evaluation of state-of-the-art FL and repair techniques, examining their advantages and limitations. Moreover, we introduce a novel approach that harnesses the power of Large Language Models (LLMs) in localising and repairing DL faults. Our evaluation, conducted on a carefully designed benchmark, reveals the strengths and weaknesses of current FL and repair techniques. We emphasise the importance of enhanced accuracy and the need for more rigorous assessment methods that employ multiple ground truth patches. Notably, LLMs exhibit remarkable performance in both FL and repair tasks. For instance, the GPT-4 model achieves 44\% and 82\% improvements in FL and repair tasks respectively, compared to the second-best tool, demonstrating the potential of LLMs in this domain. Our study sheds light on the current state of FL and repair techniques and suggests that LLMs could be a promising avenue for future advancements.
\end{abstract}

\ccsdesc[500]{Software and its engineering}

\keywords{Deep Learning, Real Faults, DL Program Repair, DL Fault Localisation}

\maketitle

\input{introduction}

\input{background}
\input{benchmark}

\input{neutrality_analysis}

\input{empirical_study_comb}

\input{results}

\input{discussion}

\input{related_work}
\input{threats_to_validity}

\input{conclusion}

\section*{Data Availability}
The data, including implementations, source code, and experimental results, are publicly available at \url{https://github.com/testingautomated-usi/dl-fl-repair}.

\bibliographystyle{ACM-Reference-Format}
\bibliography{biblio, newref, biblio2}

\newpage
\input{appendix}

\end{document}

%% file: introduction.tex
\section{Introduction}
\label{sec:intro}

Deep Learning (DL) systems are now integral to many software systems and have showcased outstanding performance across domains~\cite{Hinton2012aa,Krizhevsky2017aa,Chen2017aa,chen2015deepdriving}. As their impact grows, ensuring these models' reliability and accuracy is critical~\cite{RiccioEMSE20, sysmapHarman}. However, unlike traditional software systems, the decision logic of DL systems is not dependent only on the source code, but also on unique components such as model structure, hyperparameter selection, choice of dataset, and the underlying framework~\cite{Humbatova2020kt}. These distinctive characteristics introduce complexities and challenges when addressing faults within DL systems. Furthermore, the stochastic nature of these systems adds another layer of complexity, as retraining can lead to varying results, making it difficult to reproduce and debug issues~\cite{jahangirovantonella}.

In response, Fault Localisation (FL) and repair techniques for Deep Neural Networks (DNNs) have emerged as rapidly evolving areas within DL system testing~\cite{deepfd, wardat2022deepdiagnosis, autotrainer}. These techniques primarily focus on detecting anomalies in the training process or the model structure, which can lead to poor predictive performance of the trained model. Localising faults involves accurately pinpointing the underlying issue (e.g., incorrect loss function) according to the detected failure symptoms. Conversely, repair involves suggesting actionable fixes (e.g., changing the loss function to categorical cross-entropy). However, we believe that the reliance of these techniques on pre-defined patterns and rules may limit their effectiveness in real-world applications with diverse fault types. Furthermore, previous evaluations of FL and repair techniques have overlooked crucial aspects including the existence of multiple ground truth patches and the verification of actual improvements in model performance after applying patches to the buggy model. These gaps in previous work can lead to inaccurate assessments of the effectiveness and generalisability of FL and repair techniques.

In this paper, we present a comprehensive evaluation of FL and repair techniques for DL models, which addresses the limitations of current experimental practices. To the best of our knowledge, this is the first study that integrates five state-of-the-art FL techniques with three distinct repair strategies specifically designed for testing DL models. The FL techniques employed in our empirical study encompass a range of approaches, including the identification of problematic symptoms during training and the localisation of faults within the model structure. Simultaneously, we examine repair techniques from two disparate fields: the Software Engineering (SE) community and the Machine Learning (ML) community. This interdisciplinary exploration highlights the contrasting philosophies of the SE and ML communities: while the SE community has primarily focused on repairing DL models, the ML community has emphasised Hyperparameter Optimisation (HPO). In our evaluation, we also consider random search as a baseline repair tool, serving as a sanity check.

In addition to evaluating existing techniques, we explore the potential of employing Large Language Models (LLMs) for localising and repairing DL faults~\cite{lemieux2023codamosa, brown2020language, taxonomy}. Given that real-world faults in DL models are often a result of developer errors, these faults exhibit similar characteristics to those found in general software faults. We posit that the repetitiveness and naturalness of common faults in DL models can be effectively exploited by LLMs, as they have shown remarkable performance in FL and Automatic Program Repair (APR) for traditional software~\cite{yang2024large, xia2023automated}. We experiment with a family of GPT models from OpenAI~\cite{openai2024gpt4}, varying its temperature settings, and compare its effectiveness against existing FL and repair techniques. 

We conduct experiments on a carefully curated benchmark of faults. This benchmark comprises faults obtained through the artificial injection of defects into well-performing DL models and reproduced real-world DL faults. It includes models of varying structure and complexity, solving problems from different domains. Furthermore, we perform a neutrality analysis~\cite{neutrality} by seeking alternative patches that are equivalent or, in some cases, outperform the known ground truth patches. This analysis enhances the accuracy and robustness of our evaluation, providing a more reliable assessment of the effectiveness of FL techniques.

Our extensive evaluation with seven research questions shows that, while existing FL techniques are stable and efficient, there is considerable room for improvement in terms of FL effectiveness. Specifically, the accuracy of FL techniques is relatively low, with a maximum average recall of 0.31 and precision of 0.23 when compared to the provided ground truth. However, by extending the ground truth with our neutrality analysis, we observe a significant improvement in FL performance (maximum recall increases to 0.61), emphasising the importance of diverse ground truth variants. Interestingly, we find that LLMs demonstrate remarkable performance on FL tasks, achieving the highest performance (average recall of 0.91) in the shortest amount of time. Turning to the repair techniques, our results suggest that while current techniques can fix some faulty models, there is significant room for advancement. Interestingly, the random baseline often outperforms the most advanced repair technique from SE, and shows competitive performance compared to HPO techniques from ML. Nonetheless, none of the studied methods consistently performs well on larger and more complex models. Once again, LLMs stand out for their exceptional performance in model repair, surpassing all existing repair techniques in terms of both effectiveness and stability.

The contribution of the paper is as follows:
\begin{itemize}
    \item We conduct a comprehensive review and empirical evaluation of the current literature on FL and repair techniques for DL models. Our study focuses not only on the effectiveness of these techniques but also on their stability across multiple runs and their efficiency.
    
    \item We provide a carefully curated benchmark of repairable faults for various DL benchmark datasets and tasks. This includes both real-world faults and artificial faults using DL mutations. Additionally, we extend the ground truth fixes by performing a neutrality analysis to account for multiple alternative solutions, enhancing the evaluation process.
    
    \item We discuss when and why current FL and repair techniques fail or succeed, offering valuable insights into the factors influencing their performance. This analysis opens up new research directions for more robust and effective techniques.
\end{itemize}
    
This paper is an extended version of our two previous papers regarding FL~\cite{emp_fl} and repair~\cite{emp_repair} for DL models. We have extended them with the following technical contributions:
\begin{itemize}
    \item We explore the application of LLMs in localising and repairing DL faults by crafting a tailored prompt to effectively guide the LLMs for our specific task.
    
    \item We conduct an extensive experiment that compares LLM's performance against state-of-the-art DL FL and repair techniques to evaluate their effectiveness, efficiency, and stability, with further in-depth discussions.
    
    \item We explore alternative ground truths for repair techniques to investigate their impact on repair effectiveness and patch complexity.

    \item  We emphasise the outstanding performance of LLMs in these tasks and discuss the disparity between FL and repair in traditional SE, providing insights into potential causes for such disparity.

\end{itemize}

%% file: background.tex
\section{Background}
\label{sec:background}

This section introduces prior work on fault localisation and repair strategies for DL models. Furthermore, we discuss the potential of LLMs for these tasks.

\subsection{Automated DL Fault Localisation} \label{sec:background_fl}
Most of the proposed approaches for fault localisation for DL systems focus
on analysing the run-time behaviour during model training~\cite{deeplocalize,
wardat2022deepdiagnosis, schoop2021umlaut}. These approaches collect information and compare it to predefined rules to determine if there are any abnormalities that indicate potential faults. In the following, we provide an overview of existing FL approaches.

\textbf{DeepLocalize and DeepDiagnosis.} \DL (DL)~\cite{deeplocalize} collects performance indicators during DNN training to detect faults. It compares them with pre-defined failure symptoms and root causes from the literature, then outputs a diagnosis with the fault type, layer, phase, and iteration. The faults that it detects include the following: "Error Before/After Activation", "Error in Loss Function", "Error Backward in Weight/$\Delta$ Weight", and "Model Does Not Learn" which suggests an incorrectly selected learning rate. \DD (DD)~\cite{wardat2022deepdiagnosis} was built on the basis of \DL by expanding the list of detectable symptoms and offering targeted suggestions. It identifies ten types of faults and suggests actionable messages such as modifying loss functions or indicating improper training data. In their empirical evaluation, the authors take the randomness associated with model training into account by running each of the compared tools 5 times.

As \DL does not provide an output that can be translated into a specific fault affecting the model, we only use \DD in the empirical comparison of fault localisation tools. 

\textbf{UMLAUT.} \UM (UM)~\cite{schoop2021umlaut} combines dynamic monitoring with heuristic static checks of model structure and parameters during the training process. It includes ten heuristics from various sources, divided into "Data Preparation", "Model Architecture", and "Parameter Tuning". The output is a list of violated heuristics. The empirical evaluation of \UM was performed with 15 human participants and aimed mostly to determine whether it is useful for the developers. The authors considered  6 bugs artificially injected across two DL systems. %

\textbf{Neuralint.} Nikanjam~\etal~\cite{nikanjam2021automatic} introduced \NL (NL), a model-based fault detection approach for DL programs. It utilises meta-modelling and graph transformations to construct a comprehensive meta-model of DL programs, capturing their structure and properties. \NL then employs a set of 23 pre-defined rules, categorised into four high-level root causes~\cite{Zhang:2018}, to verify and identify potential inefficiencies in the program. These rules encompass checks for layer compatibility under the ``Unaligned Tensor'' category, optimiser and parameter initialisation under the ``API Misuse'' category, and appropriate weight/bias initialisation under the ``Incorrect Model Parameter or Structure'' category. Additionally, the ``Structure Inefficiency'' category includes rules for detecting design flaws, such as ensuring a proper decrease in neurons in fully connected layers. %

\textbf{DeepFD.} \dfd (DFD)~\cite{deepfd} is a learning-based fault detection framework for DL programs, utilising mutation testing and popular ML algorithms. It trains classifiers on a dataset of correct and faulty models, with faults injected through mutations like changing loss functions or learning rates. \dfd extracts features from runtime data and trains classifiers using K-Nearest Neighbors, Decision Tree, and Random Forest algorithms. It outputs a list of detected faults with affected code lines. The evaluation compares \dfd to \AT and \DL, accounting for stochasticity with 10 runs. %

\subsection{Automated DL Repair} \label{sec:background_repair}

Within the ML community, there has been an indirect approach to repairing model architecture through hyperparameter optimisation (HPO) techniques. While HPO techniques are primarily employed for selecting initial hyperparameters, it is also useful for enhancing under-performing models. The scope of HPO includes the optimisation of various DNN architecture components, such as layer attributes, activation functions, and even the overall network structure through adjustments to the number of layers or neurons. This perspective aligns with the SE notion of the DNN model architecture repair~\cite{emp_repair}, and thus, these techniques are encompassed in our empirical study. Within the SE community, direct approaches to model architecture repair have been proposed. Notably, \AutoT~\cite{autotrainer} introduced a method for automatically identifying symptoms in a training process and repairing them in a DNN architecture.

\textbf{Hyperparameter Optimisation. } Hyperparameter optimisation (HPO) aims to find values for hyperparameters such that the model achieves acceptable performance on a given task~\cite{claesenMoor2015hyperparameter}. With the rise of deep learning, manual HPO has become impractical, leading to automated approaches~\cite{claesenMoor2015hyperparameter, feurerHutter2019hyperparameter, zela2018towards}. Simple techniques like \textit{grid search}~\cite{montgomery2017design} have limitations: they suffer from exponential complexity with increasing hyperparameters~\cite{feurerHutter2019hyperparameter}, highlighting the need for more efficient methods. \textit{Random search}~\cite{bergstra2012random} has been proposed as a baseline, but more advanced algorithms are required to effectively navigate the complex hyperparameter space of DNNs.

\textit{Bayesian Optimisation (BO)} is a state-of-the-art strategy for global optimisation of objective functions that are costly to evaluate~\cite{feurerHutter2019hyperparameter, brochu2010tutorial}. It iteratively constructs a probabilistic surrogate model (e.g., Gaussian process) of the objective function (e.g., model accuracy given the hyperparameters) and utilises an acquisition function to balance exploration and exploitation in the hyperparameter search space~\cite{feurerHutter2019hyperparameter, Cowen-Rivers2022lm, brochu2010tutorial}. BO techniques are efficient with respect to the number of model trainings and evaluations they require~\cite{jones2001taxonomy, sasena2002flexibility}, and produced prominent results in the optimisation of DL network hyperparameters in different  domains~\cite{snoek2012practical, snoek2015scalable, melis2017state, dahl2013improving}.

\textit{HEBO (Heteroscedastic Evolutionary Bayesian Optimisation)}~\cite{Cowen-Rivers2022lm} is a state-of-the-art BO algorithm designed to optimise hyperparameters. This approach won the NeurIPS 2020 annual competition that evaluates black-box optimisation algorithms on real-world score functions. HEBO employs nonlinear transformations to handle complex noise processes and utilises multi-objective acquisition functions with evolutionary optimisers to reach a consensus among different acquisition functions. This approach allows HEBO to effectively navigate the hyperparameter search space. Another popular family of HPO approaches, called \textit{bandit-based strategies} ~\cite{feurerHutter2019hyperparameter,jamieson2016non,hyperband}, has recently been combined with BO, achieving promising results. The main representative of these combined approaches is \textit{BOHB}~\cite{bohb}.

For our evaluation, we chose \textit{Random Search} as  baseline approach and included the two state-of-the-art HPO algorithms that perform best, HEBO and BOHB.

\textbf{AutoTrainer.} \AutoT~\cite{autotrainer} is an approach that aims to detect and repair potential DL training problems. It takes as input a trained DL model saved in the ``.h5'' format and a file that contains training configurations of the model such as optimisation and loss functions, batch size, learning rate, and training dataset name.  Given a DL model and its configuration, \AutoT starts the training process and records training indicators, such as accuracy, loss values, calculated gradients for each of the neurons. It then analyses the collected values according to a set of pre-defined rules and recognises potential training problems. In its current version, the supported symptoms of training problems are: vanishing and exploding gradients, dying ReLU, oscillating loss, and slow convergence.
Once a problem has been detected, \AutoT applies its own built-in repair solutions one by one based on a default order, if an alternative, preferred order is not specified, and checks whether the problem has been fixed with the built-in solution.  The list of predefined solutions includes adding batch normalisation layers,  adding gradient clipping, adjusting batch size and learning rate, substituting activation functions, initialisers, and optimisation functions. It should be noted that when applying the possible repair solutions, \AutoT does not retrain the model with the applied repair from scratch but starts from the already trained initial model and continues the training process for more epochs with the applied solution. If none of the solutions can fix the problem, \AutoT reports its failure to find a repair to the user.

\subsection{Large Language Models (LLMs)}

LLMs have emerged as versatile, general-purpose tools with broad applicability in the tasks of Natural Language Processing (NLP) and Software Engineering (SE)~\cite{lemieux2023codamosa, brown2020language}. Models such as ChatGPT~\cite{achiam2023gpt} leverage the availability of large-size corpora of human-written text for self-supervised training (e.g., via token masking), producing trained models that can assist users on a diverse set of tasks (e.g., question-answering). With appropriate prompting, LLMs have been successfully applied to a variety of SE problems, including FL~\cite{yang2024large}, repair~\cite{xia2023automated}, and test generation~\cite{lemieux2023codamosa}. They have consistently outperformed traditional methods and this superior performance can be attributed to the predictive power of LLMs, particularly when addressing repetitive developer processes that contribute to the naturalness of the software~\cite{Hindle:2012kq}. In this paper, we explore the application of LLMs to FL and repair of DL programs, a domain that often involves developer-induced faults, which satisfies the naturalness assumption that makes LLMs effective on other SE tasks. We argue that LLMs are well-suited for localising DL faults and for improving the performance exhibited by a given DNN model architecture. In Section~\ref{sec:prompt}, we provide details on our approach to prompting LLMs for FL and repair for DL programs. For our experimental evaluation (Section~\ref{sec:empirical_study}), we use GPT-3.5, GPT-4, and GPT-4T, and study their effectiveness and efficiency in repairing faulty models compared to existing techniques, while we restricted ourselves to GPT-4 for the FL task as it required a substantial manual effort to process the output. In particular, given a prompt with a FL task for a given faulty program, the LLM of choice would return a numbered list of possible fault causes described with natural text. Processing this list and mapping it to fault types is performed manually. In addition, each prompt is queried ten times to handle the non-determinism affecting the LLM answer. 

%% file: benchmark.tex
\section{Benchmark}
\label{sec:benchmark}

To evaluate and compare FL and repair techniques selected for the study, we construct a comprehensive benchmark of faulty DNNs. Our benchmark includes two types of faults: synthetic faults, which are artificially seeded, and real faults. We further enhance these real faults through a neutrality analysis, leading to additional ground truth repairs. This carefully curated set of faults ensures a precise evaluation of the selected FL and repair techniques for DNNs. 

\begin{table}[!t]
    \caption{Fault types and their abbreviations}
    \begin{center}
    \begin{tabular}{c|l}
        \toprule
        Abbreviation & Fault Type \\ 
        \midrule
        HBS & Wrong batch size\\
        HLR & Wrong learning rate \\
        HNE & Change number of epochs \\
        ACH & Change activation function  \\
        ARM & Remove activation function \\
        AAL & Add activation function to layer \\
        RAW & Redundant weights regularisation \\
        WCI & Wrong weights initialisation \\
        LCH & Wrong loss function \\
        OCH & Wrong optimisation function \\
        LRM & Missing layer \\
        LAD & Redundant layer \\
        LCN & Wrong number of neurons in a layer \\
        LCF & Wrong filter size in a convolutional layer \\
        BCI & Wrong bias initialisation \\
        CPP & Wrong data preprocessing \\
        \bottomrule
    \end{tabular}
    \end{center}
    \label{tab:abbr}
\end{table}

\subsection{Fault Types} \label{sec:FaultTypes}

We have compiled a list of the various fault types that affect faulty models in our benchmark or are suggested in the output of the evaluated fault localisation tools, as shown in Table~\ref{tab:abbr}. The fault types are accompanied with abbreviations that will be used throughout the paper to represent specific faults and to ensure clarity and consistency. Most of the abbreviations (except the last two) are adopted from mutation operators of the DL mutation tool DeepCrime~\cite{deepcrime}, which implements mutations based on a taxonomy of real DL faults~\cite{taxonomy}. These abbreviations provide a concise way to indicate the type of fault and the affected layers. For instance, `ACH(1, 3)' signifies that the activation function needs modification in layers 1 and 3.

\subsection{Artificial Faults}

\begin{table}[ht]
\caption{Benchmark of artificial faults}
\begin{center}
\begin{tabular}{c|cccccc}
    \toprule
    Fault Type & MN & UE & CF10 & AU & UD & RT \\
    \midrule
    HLR  & M1 & U5 & - & - & - & R3 \\
    HNE  & - & U4 & C2 & A1 & - & - \\
    ACH  & - & U3 & C1 & - & - & R2 \\
    ARM  & M3 & - & - & - &- & R7 \\
    AAL  & - & U1 & - & - &- & -\\
    RAW  & - & U2 & - & - &- & R1 \\
    WCI  & M2 & U8 & C3 & - &- & R6 \\
    LCH  & - & U6 & - & A2 & D1 & R4 \\
    OCH  & - & U7 & - & - & D2 & R5 \\
    \bottomrule
    \end{tabular}
\end{center}
\label{tab:AFmodels}
\end{table}

Mutation testing is a software testing approach that involves injecting artificial faults, known as mutations or mutants, into a program~\cite{Jia:2011nx}. A test suite is considered effective if it can detect and expose these injected faults. The application of mutation testing to DNNs presents unique challenges due to the significant differences between traditional software and DNNs~\cite{jahangirovantonella}. Recently, researchers have proposed DNN-specific mutation operators that can be categorised into two main groups: pre-training and post-training mutation operators. Post-training operators~\cite{munn, deepmut, deepmut++} are applied to a trained model by modifying its structure or weights. For example, they might delete a selected layer or add Gaussian noise to a subset of weights. However, these mutations are not designed based on real-world faults and previous work~\cite{NHGJPT21} empirically showed their lower sensitivity to the changes in test set quality. Despite this limitation, post-training mutations are fast to
generate and can be preferable in settings with limited time and resources.

On the other hand, pre-training mutation operators~\cite{deepmut, NHGJPT21} inject faults directly into the source code/data of DL programs before training. This includes manipulating training data, modifying model architecture, and changing hyperparameters. Despite their huge computational cost originating from the need for re-training after mutation, these operators are more sensitive to the quality of test data~\cite{NHGJPT21}. \dc~\cite{NHGJPT21} implements some pre-training mutation operators based on a systematic analysis of real faults in DL models~\cite{Humbatova2020kt, Islam, Zhang:2018}. For our evaluation, we chose to use \dc for crafting artificial faults as it produces mutants that emulate real-world faults encountered by developers.

The replication package of \dc~\cite{deepcrimeReplication} comes with a set of pre-trained and saved mutants that cover a range of diverse DL tasks. Specifically, \dc was applied to a model for handwritten digit classification  based on the MNIST dataset~\cite{mnist} (MN), to a predictor of the eye gaze direction from an eye region image~\cite{unityeyes} (UE or UnityEyes), to a self-driving car designed for the Udacity challenge (UD), to a model that recognises the speaker from an audio recording (AU), to an image classifier for the CIFAR10 dataset~\cite{cifar10} (CF10), and to a Reuters news categorisation model~\cite{reutersdataset} (RT).

In total, the faulty model dataset of \dc consists of 850 distinct mutants. We examined all of them and selected the mutants that were killed by the test dataset provided with the subjects, according to the statistical mutation killing criterion proposed by Jahangirova and Tonella~\cite{jahangirovantonella}, which requires a statistically significant drop in prediction accuracy when the mutant is used to make predictions on the test set. In our evaluation, we adopt this statistical notion of fault exposure, with the default parameters of \dc~\cite{NHGJPT21}: $p$-value $< 0.05$ and \textit{non-negligible} effect size.

We then further filtered the mutants provided by \dc.
First of all, out of the pool of the selected mutants, we have excluded those that were generated with the help of mutation operators that affect training data, such as, for example, removing a portion of the training data or adding noise to the data, as these are not model architecture faults, hence they are out of the scope of the considered DNN FL/repair techniques. After evaluating the remaining mutants, we introduced thresholds on the performance drop to filter out  mutants that are potentially too easy to detect and repair (have a dramatic drop in performance metric when compared to the original) or those that could be too hard to repair (have a performance comparable to the original one, despite the statistical significance of the difference).
Specifically, we discarded mutants that have an average accuracy drop lower than 10\%pt\footnote{Percentage points, indicated as \%pt, is the standard unit of measure for percentage differences (e.g., a drop from 50\% to 40\% is a 10\%pt percentage points, but a 20\% percentage, drop).} of the original model's accuracy and those that are less than 15\%pt worse than the original. As for the regression systems, we kept the mutants that have an average loss value between 1.5 and 5 times of the original model's loss. %

To keep the computational costs affordable,
when more than one mutant was left after filtering for a given mutation operator, we have randomly selected one per dataset for inclusion in the final benchmark. For example, if for the `change optimisation function operator', we were left with two suitable mutants of the MNIST model, which were obtained by changing the original optimiser to either SGD or Adam~\cite{kingma2014adam}, we took only one of them randomly. After applying the described filtering procedure, we were left with 25 faulty models suitable for repair.

As a result, our benchmark contains 25 artificial DL faults across six subject models, split by nine fault types, as shown in Table~\ref{tab:AFmodels}. However, we note that our empirical evaluation excludes two artificial faults from both AU and UD, respectively, because a single experiment on them with repair techniques and Random exceeds 48 hours. Furthermore, since \DD was not applicable to SR, UD, and UE, we had to limit our fault localisation evaluation to the remaining subjects.

\subsection{Real Faults} \label{sec:real_faults}

\begin{table}[htb]
    \caption{Benchmark of real faults}
    \begin{small}
    \begin{center}
    
    \begin{tabular}{cccl}
    \toprule
    Id & SO Post \# & Task & Fault types\\
    
    \midrule
    
    D1 & 31880720 & C & ACH \\
    \midrule
    
    D2 & 41600519 & C & OCH, HBS, HNE  \\
    
    \midrule
    
    D3 &  45442843 & C & OCH, LCH, HBS, ACH, HNE   \\

    \midrule
    D4 & 48385830 & C & ACH, LCH, HLR   \\
    
    \midrule
    D5 & 48594888 & C & HNE, HBS   \\
    \midrule
    
    D6 & 50306988 & C & HLR, HNE, LCH, ACH   \\
    
    \midrule 
    D7 & 51181393 & R & HLR  \\
    \midrule
    D8 & 56380303 & C & OCH, HLR   \\
    \midrule
    
    D9 & 59325381 & C & CPP, ACH, HBS   \\

    \bottomrule
    \end{tabular}
    \end{center}
    \end{small}
    \label{tab:RFmodels}
\end{table}

To enhance our dataset of artificial faults with real faults, we leverage the benchmark of \dfd, an automated DL fault diagnosis and localisation tool~\cite{Cao2022zz}. Their original benchmark contains 58 buggy DL models collected from StackOverflow (SO) and GitHub, along with their repaired versions. We first checked if the reported faulty model, training dataset,  fault, and fault fix correspond to the original SO post or GitHub commit. We then attempted to reproduce such faults and discarded the issues where it was not possible to expose the fault in the buggy version of the model or there was no statistically significant performance improvement in the repaired version. As a result of such a filtering procedure, we were left with 9 real faults, all coming from SO. Table~\ref{tab:RFmodels} lists these faults, along with the SO post ID, task, and fault types. Of these nine faults, eight are aimed at solving a classification task (`C'), and
one is for a regression problem (`R').

%% file: neutrality_analysis.tex
\section{Alternative Ground Truth}

While our benchmark comes with a specific repair for each faulty model, there could in principle exist other hyper-parameter combinations and architectures that perform better than the faulty model and hence represent an alternative, valid fix. Consequently, we posit that there exists a potential for finding alternative patches that complement the known patch by suggesting different ways of repairing the model. Moreover, sometimes these alternative patches may possibly exhibit even better performance than the known patch. We believe that identifying such alternative patches would significantly enhance our ability to assess FL and repair techniques.

\begin{algorithm}[ht]
  \small
  \SetCommentSty{mycommfont}
  \SetKwInput{KwData}{Input}
  \SetKwInput{KwResult}{Output}

  \KwData{Initial model $s$, GT accuracy $acc_{gt}$, stopping conditions $SC$, and $top_k$}
  \KwResult{Edges $E$ and alternative GTs $R$}
    $Q, Visited, R \leftarrow \emptyset, \emptyset, \emptyset$\\
    $acc_{s} \leftarrow trainAndEvaluate(s)$\\
    $Q.enqueue([s, acc_{s}])$\\
    
    \While{$Q \neq \emptyset$ and $SC$ not met}{
      $c, acc_{c} \leftarrow Q.dequeue()$\\
      $Visited.append(c)$\\
      
      $N \leftarrow getNeighbours(c, Visited)$\\
      $tempQ \leftarrow \emptyset$ \\
      \ForEach{$n$ in $N$}{
        $acc_{n} \leftarrow trainAndEvaluate(n)$\\

        \tcp{Check whether the neighbour is equivalent to or better than the current node.}
        \If{$isNeutral(acc_{n}, acc_{c})$}{
          $tempQ.append([n, acc_{n}])$\\
        }
        \tcp{Check whether the neighbour is equivalent to or better than the given GT.}
        \If{$isNeutral(acc_{n}, acc_{gt})$}{
          $R.append([n, acc_{n}])$\\
        }
      }
      \tcp{Enqueue top $k$ neighbours to $Q$ and make edges to them.}
      $tempQ \leftarrow sort(tempQ, top_k)$\\
      \ForEach{$n, acc_{n}$ in $tempQ$}{
         $Q.enqueue([n, acc_{n}])$\\
         $E.append(c, n)$\\  
      }
    }
    
    \Return{$E, R$}\\

  \caption{Breadth-First Search (BFS) for Neutrality Analysis}
  \label{algo:bfs}
\end{algorithm}

\subsection{Neutrality Analysis via BFS}

In our search for alternative patches, we are inspired by the notion of \textit{mutation neutrality}, which states that a random mutation to an executable program is considered \emph{neutral} if the behaviour of the program on the test set does not change~\cite{neutrality}. Correspondingly, \textit{neutrality analysis} aims at finding diverse mutations with similar \textit{fitness values}, measured by a function $f$ that characterizes the program's behaviour. Whenever neutrality analysis finds an equivalent mutation pair, $\langle\mu_1$, $\mu_2\rangle$, with $f(\mu_1) \simeq f(\mu_2)$, the edge $\langle\mu_1$, $\mu_2\rangle$ is added to the  \textit{neutrality graph} (or \textit{neutrality network}~\cite{neutrality}) produced by the analysis (initially, the neutrality graph contains just the program under analysis, with no edges).

In our setting, the alternative patches available in the neutrality graph can be utilised as alternative Ground Truths (GTs). Since our targets are DL programs, the conditions for performing neutrality analysis differ from those of traditional programs. For example, the fitness is now measured by the model performance with standard metrics such as test set accuracy. This means that fitness evaluation involves training and testing of the model. Moreover, during fitness evaluation, it is important to account for the inherent stochasticity of training. To address this, in our algorithm below, we train the model ten times and calculate the fitness as an average of the resulting ten performance (e.g., accuracy) values.

Algorithm~\ref{algo:bfs} presents the Breath-First Search (BFS) for our neutrality analysis on DL programs. This algorithm takes as inputs an initial (buggy) model $s$, the accuracy of the known GT $acc_{gt}$, and stopping criteria $SC$. The outputs are a list of alternative GTs and edges of the neutrality graph. The algorithm starts with training and evaluating the initial buggy model before putting it in the queue (Lines 2-3). Next, it begins a search loop where it iteratively retrieves a model (i.e., a parent model $c$) along with its accuracy $acc_{c}$ from the queue (Line 5). Subsequently, the algorithm explores all adjacent models (i.e., neighbours) that are obtained by applying a distinct single mutation on $c$ (Line 7). Each mutation involves changing a single hyperparameter of the model, in other words, neighbouring models differ from their parent model by one hyperparameter. The mutation operators adopted from Table~\ref{tab:abbr} are HBS, HLR, HNE, LRM, LAD, LCN, ACH, BCI, WCI, LCH, OCH.
Then, the algorithm iterates over the neighbours by training and evaluating them (Line 10), and evaluates the neutrality of each neighbour compared to the parent model (Line 11) and the known GT (Line 13) (a model is considered \textit{neutral} relative to another model if it has equivalent or higher fitness than the other's, by comparing the mean accuracy of ten trained instances of the model and the other model). 
Since sometimes the number of neutral neighbours is numerous, potentially impeding the exploitation of the search, the algorithm places them into the temporal queue \textit{tempQ}, not in the main queue $Q$ (Line 12). If a node is neutral with respect to the known GT, it is added to the list $R$ of alternative GTs (Line 14). After this iteration, the algorithm sorts the temporal queue \textit{tempQ} by accuracy and takes only the top-$k$ performing neighbours by enqueueing them into the main queue $Q$. The search process stops when it meets the given stopping criteria $SC$ or the main queue $Q$ is empty. As the algorithm evolves the model by applying mutations to its own parent, which might in turn be a mutation of the original model, the resulting alternate GTs are usually higher-order mutants of the initial fixed model.

\begin{figure}[!t]
    \centering
    \includegraphics[width=0.55\linewidth]{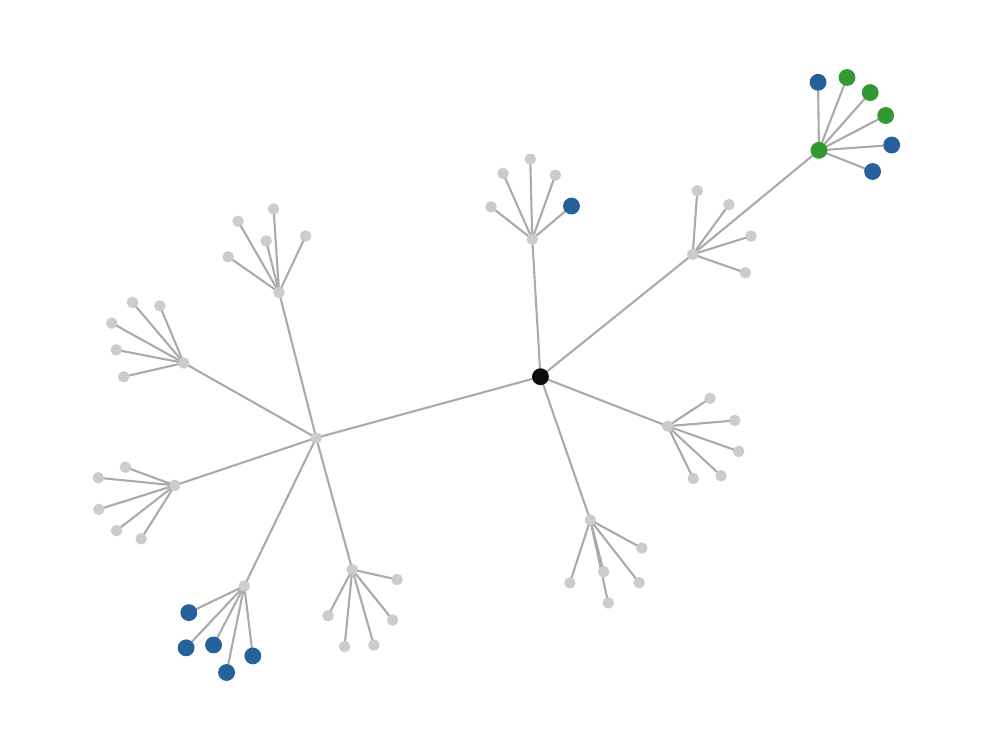}
    \caption{An example neutrality graph for the known patch (black node) of D4; green (resp. blue) nodes improve the performance of the known patch in a statistically significant (resp. insignificant) way; grey nodes are neutral to their parent}
    \label{fig:network}
\end{figure}

Based on the search results, we can draw the neutrality graph, as shown in Figure~\ref{fig:network}. Each edge represents a single mutation and each node represents the DL models (i.e., mutants). A black node denotes the initially known patch and the other nodes are neutral with their parent node. Among them, the ones that are on par with or better than the known patch are coloured in either blue or green. In particular, models that outperform the known GT with statistical significance are marked blue.\footnote{For the computation of statistical significance, we employ a Generalized Linear Model (GLM) with a significance level 0.05 and Cohen's $d$ to measure the effect size, for which we adopt a threshold at 0.5.} 
Among the other nodes, those that satisfy the neutrality condition and are equivalent to or better than the known patch (i.e., those in list $R$ returned by Algorithm~\ref{algo:bfs}, blue nodes excluded) are coloured in green. 
In this example, we found 13 alternative patches that fix the buggy model differently and show higher performance than the known patch.

\subsection{Results of Neutrality Analysis}

\begin{table}[!h]
    \centering
  \caption{Results of Neutrality Analysis: columns indicate the fault Id, the number of nodes in the neutrality graph (`\# Node'), the number of alternative GTs (`\# Alternative GTs'), the average (total) number of affected hyperparameters in the alternative GTs (`Complexity') and the average performance improvement}
  \label{tab:neutrality_stats}
    \scalebox{0.85}{
    \begin{tabular}{c|c|c|c|c}
    \toprule
    Id & \# Node  & \# Alternative GTs & Complexity & Improvement \\
    \midrule
M1 & 258 & 240 & 3.54 (21) & 0.000 \\
M2 & 291 & 291 & 2.54 (21) & 0.000 \\
M3 & 170 & 170 & 2.01 (21) & 0.000 \\
C1 & 61 & 36 & 1.86 (27) & 0.003 \\
C2 & 31 & 1 & 1.00 (27) & 0.007 \\
C3 & 19 & 10 & 2.00 (27) & 0.004 \\
R1 & 45 & 0 & - (12) & - \\
R2 & 57 & 55 & 1.98 (12) & 0.008 \\
R3 & 60 & 0 & - (12) & - \\
R4 & 19 & 19 & 1.68 (12) & 0.009 \\
R5 & 31 & 19 & 2.58 (12) & 0.008 \\
R6 & 23 & 20 & 1.90 (12) & 0.004 \\
R7 & 38 & 38 & 2.79 (12) & 0.008 \\
\midrule
\textbf{Avg.} &\textbf{84.85} & \textbf{69.15} & \textbf{2.17 (18.55)} & \textbf{0.00} \\
\midrule
D1 & 92 & 92 & 4.29 (15) & 0.000 \\
D2 & 14 & 0 & - (21) & - \\
D3 & 47 & 44 & 8.59 (13) & 0.003 \\
D4 & 61 & 13 & 9.54 (12) & 0.010 \\
D5 & 41 & 1 & 4.00 (19) & 0.001 \\
D6 & 37 & 7 & 5.29 (12) & 0.000 \\
D7 & 49 & 25 & 4.04 (9) & 0.065 \\
D8 & 73 & 73 & 8.51 (17) & 0.186 \\
D9 & 22 & 0 & - (19) & - \\
\midrule
\textbf{Avg.} &\textbf{48.44} & \textbf{28.33} & \textbf{6.32 (13.86)} & \textbf{0.04} \\

            \bottomrule
    \end{tabular}
    }
\end{table}

Table~\ref{tab:neutrality_stats} presents the results of neutrality analysis. Column `\# Node' shows the number of nodes in the neutrality graph, where each node is \textit{neutral} with respect to its parent, along with the count of alternative GTs (`\# Alternative GTs')  that match or exceed the performance of the original GT, as illustrated by the green and blue nodes in Figure~\ref{fig:network}. The `Complexity' column reflects the degree of variation between each alternative and the known GT by counting the differing hyperparameters; for each row, complexity is averaged across all identified alternatives. The number in parentheses indicates the total number of hyperparameters mutated in the alternative GTs per fault. The `Improvement' column shows the extent of performance enhancement over the known GT, based on the chosen performance metric, calculated as the average difference across all identified alternative GTs. For faults R1, R3, D2, and D9, no alternative patches were found within the set budget, so no results are reported (marked as \verb|-|).

Our neutrality analysis uncovers an average of 69 alternative GTs for artificial faults and 28 for real faults, suggesting the potential influence of alternative GTs on the evaluation of FL tools. Typically, real faults show greater complexity, with an average of 6.32 affected hyperparameters, compared to 2.17 for artificial faults. This difference likely arises because artificial faults are intentionally simpler, involving only one hyperparameter mutation relative to the GT, whereas real faults tend to be more complex. For what concerns the magnitude of performance improvement by alternative GTs over the known GT, we note only modest gains, which are slightly more pronounced for real faults than for artificial ones. In the subsequent section (Section~\ref{sec:empirical_study}), we will integrate these newly discovered GTs into our fault dataset to investigate their impact on FL effectiveness and patch complexity analysis.

%% file: empirical_study_comb.tex
\section{Empirical study}
\label{sec:empirical_study}

\subsection{Research Questions}

The main \textit{goal} of our empirical study is to compare existing approaches for DL fault localisation and repair, among each other and with the capabilities of LLMs (e.g., GPT-4~\cite{openai2024gpt4}) prompted for these two tasks. The comparison is conducted on our benchmark of artificial and real faults (see Section~\ref{sec:benchmark}). We design our study to investigate the following seven research questions:

\begin{itemize}
    \item \textbf{RQ1. FL Effectiveness}: \textit{Can existing FL tools locate faults correctly in DL models? Are LLMs more effective in this task? How do the outcomes differ when considering alternative GTs?}
    
    \item \textbf{RQ2. FL Stability}: \textit{Is the outcome of  FL tools stable across multiple runs?}
    
    \item \textbf{RQ3. FL Efficiency}: \textit{How costly are FL tools when compared to each other and to GPT querying?}
    
    \item \textbf{RQ4. Repair Effectiveness}: \textit{Can existing DL repair tools generate patches that improve the evaluation metric? Can LLMs serve as a repair tool? Which repair tool produces the most effective patches?}
    
    \item \textbf{RQ5. Repair Stability}: \textit{Are the patches generated by existing DL repair tools stable across several runs? How does the temperature parameter influence the stability of LLMs?}
    
    \item \textbf{RQ6. Repair Efficiency}: \textit{How much does the performance of the repair tools change when having a smaller or bigger budget?}
    
    \item \textbf{RQ7. Patch Complexity}: \textit{How complex are the generated patches? Do they align with either the original or alternative GTs?} 
\end{itemize}

RQ1 and RQ4 are the key research questions for our empirical study as they compare the effectiveness of different FL and repair tools to the performance of LLMs on our curated benchmark. RQ1 is further divided into two sub-questions to address the alternative GTs: either we consider just the original GT (RQ1.1) or also all alternative GTs (RQ1.2). As Table~\ref{tab:neutrality_stats} shows, the performance improvements (i.e., Column `Improvement') between the alternative and original GTs are marginal. Given that RQ4 solely evaluates the magnitude of performance improvement, regardless of its source, this RQ does not require sub-questions to differentiate between the original and alternative GTs.

RQs 2, 3, 5, and 6 investigate important properties of any FL or repair tool: its stability across multiple executions and the dependency of its outcome on the execution budget. Finally, RQ7 explores the complexity of repair patches generated by different tools and their similarity to the original GT.

\subsection{Prompts for GPTs}
\label{sec:prompt}

We designed a basic prompt for a GPT to facilitate fault localisation in the given model under test, as shown in Listing~\ref{lst:fl_prompt}. To provide GPT with some context, we specified the dataset used for training and the associated task. For less-known datasets, we omitted this detail. Additionally, we provided a general hint about fault types that may occur in a DL program, such as incorrect design or hyperparameter selection, and their possible symptoms like an underperforming model. We avoided few-shot prompting to prevent biasing GPT towards the specific fault types provided in the examples. Instead, we followed best practices for prompt engineering~\cite{gpt_prompting}. We instructed the model to present the localised faults in an ordered manner, but we did not consider this ordering when calculating the main metrics because the tools we compared with do not have such functionality.

\begin{lstlisting}[label={lst:fl_prompt}, caption={Prompt template for FL}]
The code below, delimited by triple backticks, is designed for a {task} trained on {dataset}. There may be a number of faults in this code, such as incorrect neural network design or hyperparameter selection, that cause the trained neural network to underperform. Please review the code and decide whether or not there are faults that cause this neural network to underperform when it is trained. Then provide the main reasons for the decision numbered in decreasing order of importance (from most important to least).
Code:
```{code}```
\end{lstlisting}\leavevmode

Listing~\ref{lst:repair_prompt} presents an instruction for GPT to repair some DL code by modifying its hyperparameters. GPT is tasked with determining the appropriate hyperparameters in the form of a \texttt{config} dictionary within the code, given the repair goal, the model task and dataset, and the user-provided code. Note that this prompt is designed to function without knowledge of the faults. In our preliminary study, we initially supplied GPT with a prompt pointing to the buggy code with faulty hyperparameters, but GPT did not appear to utilise this information, resulting in largely unchanged outcomes.

\begin{lstlisting}[float, label={lst:repair_prompt}, caption={Prompt template for repair}]
The following code is designed for a {task} trained on {dataset}. Please repair it in order to {goal}. The code repair consists of replacing one or more of the hyperparameters with an alternative value, currently represented in a "config" dictionary, in the form of config["PARAM"]. Please only show me config values in a JSON format so that I can save it directly in a json file format. Give me only one solution.
Code: {code}
\end{lstlisting}\leavevmode

\subsection{Experimental Settings}
\label{sec:exp_fl}

In this section, we discuss the experimental settings and evaluation metrics that we adopted to perform our empirical study.

\subsubsection{Selected Repair Operators}
\label{sec:top_repair_ops}

While the number of possible repair combinations grows exponentially with the number of hyperparameters that can be changed by the repair tools, not all repair operators are equally likely to be effective and useful in practice. To identify which hyperparameters should be given high priority while searching for a DL repair operator, we analyse the taxonomy of real faults in DL systems~\cite{Humbatova2020kt}. 
Specifically, we consider the number of issues coming from SO, GitHub and interviews that contributed to each leaf of the taxonomy and grouped similar fault types together. Given the resulting list of fault types sorted by prevalence, we only consider the top 12 entries for the purposes of this study. We excluded fault types that would typically lead to a crash, as they are out of scope when considering model architecture faults. For example, we exclude fault types related to wrong input or output shapes of a layer. This leaves us with the 12 most frequent faults. The selected fault categories include: change loss function (LCH), add/delete a layer (LRM and LAD), enable batching/change batch size (HBS), change the number of neurons in a layer (LCN), change learning rate (HLR), change number of epochs (HNE), change/add/remove activation function (ACH, ARM, and AAL), change weights initialisation (WCI), and change optimisation function (OCH). Random, HPO techniques, and GPTs are designed to repair only these 12 fault types. Note that \AutoT, by design, has a narrower focus and can only address HLR, ACH, AAL, WCI, and OCH.

\subsubsection{Processing tool output}
\label{sec:processing_output}

For FL tools, the output format varies across techniques. Once executed, \dfd provides a list of fault types detected in the given DL program. Moreover, it can specify line numbers for localised faults. In contrast, \UM generates warnings and critical messages at the end of each training epoch, typically consisting of a few words or a sentence. \DD, which utilises a tool-specific callback to monitor the training process like \UM, terminates the training and writes the identified faults into a file once any symptom is detected. These symptoms are localised to specific layers, and the tool provides corresponding fault types and potential fixes when possible. The output is usually concise, with a maximum of one short sentence per component (e.g., symptom, fault type, fix). \NL can associate faults with specific layers or a general `Learner' function and presents the identified faults with 1-2 detailed sentences. Similarly, GPT-4 generates a list of 5-10 answers (in our experiments), each consisting of 2-3 sentences. Mapping \NL and GPT-4's outputs to fault types is more effort-consuming compared to other FL tools, as the former requires manual analysis of each of the answers where fault types are not mentioned explicitly. %
In our experiments, one of the authors
analysed all outputs and provided mappings to available fault types, adding new types as necessary. For completeness, such detailed mappings are presented in Tables~\ref{tab:main_results_fl_1_4_1} - \ref{tab:main_results_fl_openai} in the Appendix (Section~\ref{sec:RQ1_appendix}).

For repair tools, we do not require further processing of the output of the HPO techniques, since they result in a model with modified hyperparameters. Regarding GPTs, we instructed it to produce the recommended parameters in JSON format, allowing us to utilise the output directly.

\subsubsection{Implementation}

For the comparison between FL tools, we adopt publicly available versions of all considered tools~\cite{deepfd_replication, umlaut_replication, neuralint_replication, deepdiagnosis_replication} that are run on Python with library versions specified in the requirements for each tool. However, we had to limit the artificial faults to those obtained using CF10, MN, and RT as \DD is not applicable to other subjects. The authors of \dfd adopted the notion of statistical mutation killing~\cite{JahangirovaICST20} in their tool. They run each of the models used to train the classifier as well as the model under test 20 times to collect the run-time features. For FL using \dfd, we adopt an ensemble of already trained classifiers provided in the tool's replication package. Similar to the authors of \dfd, for each faulty model in our benchmark, we collect the run-time behavioural features from 20 re-trainings of the model. \NL is based on static checks that do not require any training and thus, are not prone to randomness. We run each of the remaining tools 20 times to account for the randomness in the training process and report the most frequently observed result (mode).

For repair tools, we use the Ray Tune~\cite{liaw2018tune} library to implement the Random baseline, as well as HEBO and BOHB. We set the 12 chosen repair operators as the hyperparameter search space, and we change the time budget to simulate different experimental settings. Except for Random, the two HPO techniques start the search from the initial configuration of the faulty model. 

We use a publicly available version of \AutoT.\footnote{\url{https://github.com/shiningrain/AUTOTRAINER}} Our goal was to apply \AutoT to all of our subject systems. However, its current implementation does not support regression systems. As a result, \AutoT is applicable to 13 artificial faults out of 21 and eight real faults out of nine. 
As the performance of repair tools can be highly affected by the time budget, we run all experiments on three different time budgets, 10, 20, and 50, which are the multipliers of the training time of the initial faulty model.
We run each tool ten times to handle the randomness of the search and the training process, and report the average of the results. In addition, we split the test set into two parts: one for guiding the search (i.e., only used during the search to evaluate candidate patches) and the other for the final evaluation
of the generated patches at the end of the search. Note that \AutoT operates differently from HPO techniques: it only begins the repair once it diagnoses a failure symptom and continues until it does not observe any. This makes it challenging to apply the same time budget configurations as for the other HPO techniques. Instead, we simply execute \AutoT repeatedly until the total execution time reaches the maximum budget, and collect results for lower budgets by looking at the executions completed within each lower time budget. %

\subsubsection{Statistical Tests \& Evaluation Metrics}

For the FL task, we employ standard information retrieval metrics to calculate the similarity between the ground truth and the fault localisation results. These metrics include Precision (PR), Recall (RC), and $F_{\beta}$ score:

\begin{equation}
    RC = \frac{| FT_{loc} \cap FT_{gt} |}{| FT_{gt} |}
\end{equation}

\begin{equation}
    PR = \frac{| FT_{loc} \cap FT_{gt} |}{| FT_{loc} |}
\end{equation}

\begin{equation}
    F_{\beta} = (1 + \beta^2) \frac{PR \cdot RC}{\beta^2 PR + RC}
\end{equation}

Recall measures the proportion of correctly reported fault types in the list of localised faults ($FT_{loc}$) among those in the ground truth ($FT_{gt}$); Precision measures the proportion of correctly reported fault types among the localised ones; $F_{\beta}$ is a weighted geometric average of $PR$ and $RC$, with the weight $\beta$ deciding on the relative importance between $RC$ and $PR$. Specifically, we adopt $F_{\beta}$ with $\beta$ equals 3, which gives three times more importance to recall than to precision. This choice of $\beta$ is based on the assumption that in the task of fault localisation, the ability of the tool to find as many correct fault sources as possible is more important than the precision of the answer. For neutrality analysis, we set $top_k$ to 5 and the stopping condition $SC$ to a 48-hour time budget. During the search, every model is trained ten times and we use a mean of the ten metric values depending on the task solved by each subjects (i.e., accuracy for classification or loss for regression).

For the repair task, we utilise a non-parametric Wilcoxon-signed rank test to determine the statistical significance of patch improvements on the performance metric values. The null hypothesis states that the medians of two lists of metric values (from the faulty model and the patch) are equal, while the alternative hypothesis suggests they are different. We set the significance level at 0.05 to reject the null hypothesis. Furthermore, we use the following metric, named Improvement Rate (IR), to measure how much the evaluation metric of the fault ($M_{Fault}$) has been improved by the patch, in comparison with the ground truth improvement:

\begin{equation}
    IR = \frac{M_{Patch} - M_{Fault}}{M_{GT} - M_{Fault}}
\end{equation}

\noindent
where $M_{Patch}$ is the evaluation metric of the patch generated by the repair tools and $M_{GT}$ is the evaluation metric of the ground truth  model, either provided by developers (real faults) or obtained as the model before mutation (artificial faults). For example, if IR is 1, the generated patch is as effective as the ground truth fix (it can be noticed that, in principle, IR can be even greater than 1). We reverse the sign of IR when dealing with mean squared error, as lower values are better.

To quantify the stability of each DL repair tool, we measure the standard deviation $\sigma$ of the repaired model performance achieved in ten runs of each tool. 

The complexity of a patch is computed as the number of hyperparameters that differ between the generated patch and the initial faulty model. For example, if the patch only changes the batch size from 8 to 32, while all remaining hyperparameters are unchanged, the patch is considered to have a complexity of 1. We normalise the complexity metric by dividing it with the total number of hyperparameters, so that it ranges between 0 (i.e., it has the same hyperparameters as the initial faulty model)
and 1 (i.e., all hyperparameters have been changed).

Lastly, to quantify the similarity between the sets of repair operators used by the generated patch and the ground truth, we adopt the Asymmetric Jaccard (AJ) metric, which measures the percentage of ground truth repair operators ($OP_{GT}$) that also appear in the patch ($OP_{Patch}$):
\begin{equation}
    AJ = \frac{| OP_{Patch} \cap OP_{GT} |}{| OP_{GT} |}
\end{equation}

%% file: results.tex
\section{Results}
\label{sec:results}

\subsection{RQ1.1 (FL Effectiveness Before Neutrality Analysis)}
\label{sec:RQ11}

\begin{table}[!h]
  \centering
  \caption{Number of Ground Truth (GT) faults (\#F); Recall (RC), Precision (PR) and $F_3$ measure for each FL tool. Avg. shows the average within artificial or real faults. T.A. shows the total average across faults.} 
  \label{tab:main_results_fl_2}
   \scalebox{0.8}{
  \begin{tabular}{l|c|ccc|ccc|ccc|ccc|ccc}
    \toprule
   Id & GT &\multicolumn{3}{c|}{DFD}&\multicolumn{3}{c|}{DD}&\multicolumn{3}{c|}{NL}&\multicolumn{3}{c|}{UM} & \multicolumn{3}{c}{GPT-4}\\
     & \#F & RC  & PR & $F_3$ & RC  & PR & $F_3$ & RC & PR & $F_3$ & RC & PR & $F_3$  & RC & PR &$F_3$  \\
    \midrule    
M1 & 1 &0  & 0&0& 0  & 0&0& 1  & 1&1& 0   & 0&0 & 1& 0.28&0.79
\\
M2 & 1 &0  & 0&0& 1  & 1&1& 0  & 0&0& 1   & 0.5&0.91 & 1& 0.32&0.82\\
M3 & 1 &1  & 0.33&0.83& 0  & 0&0& 0  & 0&0& 0   & 0&0 & 1& 0.26&0.77\\
C1 & 1 &1  & 0.25&0.77& 0  & 0&0& 0  & 0&0& 0   & 0&0 & 1& 0.63&0.92
\\
C2 & 1 &0  & 0&0& 0  & 0&0& 0  & 0&0& 0   & 0&0 & 1& 0.80&0.96
\\
C3 & 1 &0  & 0&0& 0  & 0&0& 1  & 1&1& 0   & 0&0 & 1& 0.55&0.89\\
R1 & 1 &0  & 0&0& 0  & 0&0& 0  & 0&0& 0   & 0&0 & 1& 0.26&0.77
\\
R2 & 1 &0  & 0&0& 1  & 1&1& 0  & 0&0& 1   & 1&1 & 1& 0.25&0.76
\\
R3 & 1 &0  & 0&0& 0  & 0&0& 0  & 0&0& 1   & 1&1 & 1& 0.22&0.73\\
R4 & 1 &1  & 0.50&0.91& 0  & 0&0& 1  & 1&1& 0   & 0&0 & 1& 0.21&0.72
\\
R5 & 1 &1  & 0.33&0.83& 0  & 0&0& 0  & 0&0& 0   & 0&0 & 1& 0.20&0.71\\
R6 & 1 &0  & 0&0& 0  & 0&0& 1  & 1&1& 0   & 0&0 & 1& 0.25&0.77
\\
R7 & 1 &0  & 0&0& 1  & 1&1& 0  & 0&0& 1   & 1&1 & 1& 0.50&0.91\\
    \midrule
\textbf{Avg.} & \textbf{1} &\textbf{0.31}  & \textbf{0.11} & \textbf{0.26} & \textbf{0.23}  & \textbf{0.23} & \textbf{0.23} & \textbf{0.31}  & \textbf{0.31} & \textbf{0.31} & \textbf{0.31}  & \textbf{0.27} & \textbf{0.30} & \textbf{1}& \textbf{0.36}&\textbf{0.81}\\
    \midrule
D1 & 1 &1  & 1&1& 0  & 0&0& 0  & 0&0& 0   & 0&0 & 1& 0.50&0.90
\\
D2 & 3 &0  & 0&0& 0  & 0&0& 0  & 0&0& 0   & 0&0 & 0.77& 0.50&0.73
\\
D3 & 5 &0.2  & 0.50&0.21& 0  & 0&0& 0.4  & 0.67&0.42& 0   & 0&0 & 0.76& 0.79&0.76
\\
D4 & 3 &0  & 0&0& 0.33  & 1&0.35& 0.33  & 0.5&0.34& 0.67   & 1&0.69 & 1& 0.45&0.89
\\
D5 & 2 &0  & 0&0& 0  & 0&0& 0  & 0&0& 0   & 0&0 & 1& 0.50&0.91
\\
D6 & 4 &0.5  & 0.67&0.51& 0  & 0&0& 0  & 0&0& 0   & 0&0 & 0.35& 0.47&0.36
\\
D7 & 1 &0  & 0&0& 0  & 0&0& 0  & 0&0& 0   & 0&0 & 0.10& 0.03&0.08
\\
D8 & 2 &1  & 0.50&0.91& 0  & 0&0& 0  & 0&0& 0   & 0&0 & 0.20& 0.07&0.17
\\
D9 & 3 &0  & 0&0& 0  & 0&0& 0  & 0&0& 0   & 0&0 & 0.40& 0.49&0.40\\
    \midrule
\textbf{Avg.} & \textbf{2.67} &\textbf{0.30}  & \textbf{0.30}& \textbf{0.29}& \textbf{0.04}  & \textbf{0.11} & \textbf{0.04}& \textbf{0.08}  & \textbf{0.13} & \textbf{0.08} & \textbf{0.07}  & \textbf{0.11} & \textbf{0.08} & \textbf{0.62}& \textbf{0.42}&\textbf{0.58}\\ 			
    \midrule
\textbf{T.A.} & \textbf{1.68} & \textbf{0.31}  & \textbf{0.19} & \textbf{0.27} & \textbf{0.15}  & \textbf{0.18} &\textbf{0.15} & \textbf{0.22}  & \textbf{0.23} & \textbf{0.22} & \textbf{0.21}  & \textbf{0.20} & \textbf{0.21} & \textbf{0.84}& \textbf{0.39}&\textbf{0.71}\\
    \bottomrule
  \end{tabular}
  }
\end{table}

Table~\ref{tab:main_results_fl_2} summarises the overall assessment of the effectiveness of the FL tools.\footnote{For completeness, we report a detailed result table for each tool in Appendix (Section~\ref{sec:RQ1_appendix}).} The column `$GT \# F$' indicates the number of fault types in the ground truth, while the columns \textit{`<tool\_name>'} present all the performance metrics measured for each tool: `$RC$' sub-columns display Recall values, `$PR$' sub-columns show Precision, and `$F_3$' sub-columns represent the $F_\beta$ score with $\beta = 3$. 
To facilitate comparisons among the tools, we provide average values for each tool across both artificial and real faults (shown in the `Avg.' rows) and across all benchmark faults (the `T.A' row).

 According to all the considered metrics, GPT-4  significantly surpasses all competitors. For example, on artificially seeded faults it reaches an average recall of 1, while the highest result achieved by other tools is 0.31. Similarly, on the real-world faults, where \DD gets a recall of 0.3 and other tools less than 0.1, GPT-4's performance goes as high as 0.62. Despite the high number of fault suggestions of GPT-4, its precision values on average outperform those of other tools on both types of faults. On artificial faults, the advantage of GPT-4 on precision is just 0.05, but on the real faults it goes up to 0.12. This shows that thanks to their training on an enormous amount of DL code examples, LLMs could more easily spot problems in our FL benchmark than tools that rely on observed patterns in variables capturing the evolution of the training process (\DD, \dfd, \UM) or tools that use a set of predefined rules and best-practices (\NL and \UM). Among these tools, however, we can see that \dfd gets notably higher performance on real-world faults, and a comparable one on artificial faults.

It is important to note that, unlike the other tools, \dfd does not offer suggestions for layer indexing. This limitation means it is unclear whether a correctly detected `ACH' fault type points to the correct layer to be repaired. This situation applies to 2 out of 22 faults, and if we exclude these from the calculation of average RC, \dfd's result decreases from 0.31 to 0.21, placing it on par with \NL and \UM. Assuming \dfd identifies the correct layer with a 50\% probability (i.e., the suggested layer is either accurate or not), the mean RC value would fall down to 0.26. Additionally, for some fault types, other tools—unlike \dfd—provide more targeted suggestions, such as specific activation functions (DD, UM, GPT-4), weight initialization (NL, GPT-4), or the direction of change for the learning rate (UM, GPT-4).

\begin{tcolorbox}[boxrule=0pt,frame hidden,sharp corners,enhanced,borderline north={1pt}{0pt}{black},borderline south={1pt}{0pt}{black},boxsep=2pt,left=2pt,right=2pt,top=2.5pt,bottom=2pt]
\textbf{Answer to RQ1.1 (FL Effectiveness Before Neutrality Analysis)}: Our evaluation reveals that all FL tools, except GPT-4, display relatively low RC before neutrality analysis, as they often fail to identify faults affecting the model according to the available ground truth. GPT-4, on average, outperforms the others across all metrics, while \DD shows the lowest performance. \dfd, \NL, and \UM perform similarly on artificial faults, but \dfd achieves better results on real-world faults.
\end{tcolorbox}

\subsection{RQ1.2 (FL Effectiveness After Neutrality Analysis)} \label{sec:RQ12}

\begin{table}[!h]
  \centering
  \caption{Recall (RC), Precision (PR) and $F_3$ measure for each FL tool after neutrality analysis. Avg. shows the average within artificial or real faults. T.A. shows the total average across faults. The values that increased or decreased in comparison with the initial results (before neutrality analysis) are boldfaced or underlined, respectively. The faults for which neutrality analysis was not able to find any alternative GT are greyed out.} 
  \label{tab:neutrality_results_diff}
   \scalebox{0.87}{
  \begin{tabular}{l|ccc|ccc|ccc|ccc|ccc}
    \toprule
   Id &\multicolumn{3}{c|}{DFD}&\multicolumn{3}{c|}{DD}&\multicolumn{3}{c|}{NL}&\multicolumn{3}{c|}{UM} & \multicolumn{3}{c}{GPT-4}\\
     & RC  & PR & $F_3$ & RC  & PR & $F_3$ & RC & PR & $F_3$ & RC & PR & $F_3$  & RC & PR &$F_3$  \\
    \midrule    
M1&\textbf{0.67}& \textbf{0.5}&\textbf{0.65}& \textbf{0.5}& \textbf{1}&\textbf{0.53}& 1& 1&1.00& \textbf{0.5}& \textbf{1}&\textbf{0.53} & 1& 0.28&0.79
\\
M2&\textbf{0.67}& \textbf{0.67}&\textbf{0.67}& 1& 1&1.00& \textbf{0.5}& \textbf{1}&\textbf{0.53}& 1& 0.5&0.91 & 1& 0.32&0.82
\\
M3&1& \textbf{0.47}&\textbf{0.88}& 0& 0&0& 0& 0&0& 0& 0&0 & 1& 0.26&0.77\\
C1&1& \textbf{0.39}&\textbf{0.85}& 0& 0&0& 0& 0&0& 0& 0&0 & 1& \textbf{0.64}&\textbf{0.93}
\\
C2&0& 0&0& 0& 0&0& 0& 0&0& 0& 0&0 & 1& \underline{0.73}&\underline{0.91}
\\
C3&\textbf{1}& \textbf{0.25}&\textbf{0.77}& 0& 0&0& 1& 1&1& 0& 0&0 & 1& \textbf{0.60}&\textbf{0.90}\\
\rowcolor{lightgray}
R1&0& 0&0& 0& 0&0& 0& 0&0& 0& 0&0 & 1& 0.26&0.77
\\
R2&\textbf{1}& \textbf{0.56}&\textbf{0.91}& 1& 1&1& \textbf{1}& \textbf{1}&\textbf{1}& 1& 1&1 & 1& \textbf{0.51}&\textbf{0.89}
\\
\rowcolor{lightgray}
R3&0& 0&0& 0& 0&0& 0& 0&0& 1& 1&1 & 1& 0.22&0.73\\
R4&1& \textbf{0.57}&\textbf{0.92}& 0& 0&0& 1& 1&1& 0& 0&0 & 1& \textbf{0.50}&\textbf{0.88}
\\
R5&1& \textbf{0.5}&\textbf{0.89}& 0& 0&0& 0& 0&0& 0& 0&0 & 1& \textbf{0.42}&\textbf{0.83}
\\
R6&\textbf{0.5}& \textbf{0.25}&\textbf{0.45}& 0& 0&0& 1& 1&1& 0& 0&0 & 1& 0.25&0.77
\\
R7&\textbf{1}& \textbf{0.5}&\textbf{0.89}& 1& 1&1& \textbf{1}& \textbf{1}&\textbf{1}& 1& 1&1 & 1& 0.50&0.91\\
 \textbf{Avg.}& \textbf{0.68}& \textbf{0.36}& \textbf{0.61}& \textbf{0.27}& \textbf{0.28}& \textbf{0.27}& \textbf{0.50}& \textbf{0.54}& \textbf{0.50}& \textbf{0.35}& \textbf{0.35}& \textbf{0.34} & \textbf{1.00}& \textbf{0.42}&\textbf{0.84}\\
    \midrule
D1&1& 1&1& \textbf{0.5}& \textbf{1}&\textbf{0.53}& \textbf{0.5}& \textbf{1}&\textbf{0.53}& 0& 0&0 & 1& \textbf{0.72}&\textbf{0.95}
\\
\rowcolor{lightgray}
D2&0& 0&0& 0& 0&0& 0& 0&0& 0& 0&0 & 0.77& 0.50&0.73
\\
D3&\textbf{1}& 0.5&\textbf{0.91}& 0& 0&0& \textbf{0.5}& \underline{0.33}&\textbf{0.48}& 0& 0&0 & \textbf{0.96}& \underline{0.53}&\textbf{0.88}
\\
D4&0& 0&0& 0.33& 1&0.35& \textbf{1}& 0.5&\textbf{0.91}& \textbf{1}& \underline{0.5}&\textbf{0.91} & 1& \underline{0.31}&\underline{0.81}
\\
D5&0& 0&0& 0& 0&0& 0& 0&0& 0& 0&0 & 1& 0.50&0.91
\\
D6& \textbf{1} & \underline{0.5} &\textbf{0.89}& 0& 0&0& 0& 0&0& 0& 0&0 & \textbf{0.70}& \textbf{0.90}&\textbf{0.70}
\\
D7&\textbf{0.5}& \textbf{1}&\textbf{0.53}& 0& 0&0& 0& 0&0& 0& 0&0 & \textbf{0.55}& \textbf{0.34}&\textbf{0.50}
\\
D8&1& 0.5&0.91& 0& 0&0& 0& 0&0& 0& 0&0 & \textbf{0.70}& \textbf{0.27}&\textbf{0.53}
\\
\rowcolor{lightgray}
D9&0& 0&0& 0& 0&0& 0& 0&0& 0& 0&0 & 0.40& 0.49&0.40\\
 \textbf{Avg.}& \textbf{0.50}& \textbf{0.39}& \textbf{0.47}& \textbf{0.09}& \textbf{0.22}& \textbf{0.10}& \textbf{0.22}& \textbf{0.20}& \textbf{0.21}& \textbf{0.11}& \textbf{0.06}& \textbf{0.10} & \textbf{0.79}& \textbf{0.51} & \textbf{0.71}\\
 \midrule
 \textbf{T.A.}& \textbf{0.61}& \textbf{0.37}& \textbf{0.55}& \textbf{0.20}& \textbf{0.26}& \textbf{0.20}& \textbf{0.39}& \textbf{0.40}& \textbf{0.38}& \textbf{0.25}& \textbf{0.23}& \textbf{0.24} & \textbf{0.91}& \textbf{0.46}&\textbf{0.79}\\
    \bottomrule
  \end{tabular}
  }
\end{table}

We examined the hypothesis that relying on a single GT, defined by a single set of modifications that improve the model performance, may be inadequate for evaluating the effectiveness of FL tools. In this research question, we analyse how the evaluation metrics of various FL tools change when they are provided with a finite set of possible alternative GTs.

Following the neutrality analysis, we recalculated the evaluation metrics for each tool, incorporating all available GT variants. Table~\ref{tab:neutrality_results_diff} presents the updated results. For faults where no alternative GTs were identified, results are shaded in grey. Improved RC, PR, and $F_3$ scores following neutrality analysis appear in bold, while decreases in PR and $F_3$ due to alternative GTs are underlined. This table shows the highest average RC achieved across all GT variants, along with the average PR and $F_3$ values calculated for the corresponding GTs. This means that we match the output of each tool with the most similar GT among the available ones.

\begin{table}[!h]
  \centering
  \caption{Overall comparison of Recall (RC), Precision (PR) and $F_3$ measure for each FL tool before/after neutrality analysis. Avg. shows the average within artificial (AF) or real (RF) faults. T.A. shows the total average across faults.} 
  \label{tab:neutrality_results}
   \scalebox{0.8}{
  \begin{tabular}{l|ccc|ccc|ccc|ccc|ccc}
    \toprule
   Id &\multicolumn{3}{c|}{DFD}&\multicolumn{3}{c|}{DD}&\multicolumn{3}{c|}{NL}&\multicolumn{3}{c}{UM} &\multicolumn{3}{c}{GPT-4}\\
     & RC  & PR & $F_3$ & RC  & PR & $F_3$ & RC & PR & $F_3$ & RC & PR & $F_3$ & RC & PR & $F_3$\\
    \midrule
 \multicolumn{13}{c}{Before neutrality analysis}\\
 \midrule
 \textbf{AF Avg.}& 0.31& 0.11& 0.26& 0.23& 0.23& 0.23& 0.31& 0.31& 0.31& 0.31& 0.27&0.3 & 1 & 0.36 & 0.81\\
 \textbf{RF Avg.}& 0.3& 0.3& 0.29& 0.04& 0.11& 0.04& 0.08& 0.13& 0.08& 0.07& 0.11&0.08 & 0.62 & 0.42 & 0.58\\
 \textbf{T.A.} & 0.31& 0.19& 0.27& 0.15& 0.18& 0.15& 0.22& 0.23& 0.22& 0.21& 0.2&0.21 & 0.84 & 0.39 & 0.71 \\
 \midrule
 \multicolumn{13}{c}{After neutrality analysis}\\
 \midrule
\textbf{AF Avg.}&0.68& 0.36& 0.61& 0.27& 0.28& 0.27& 0.50& 0.54& 0.50& 0.35& 0.35& 0.34 & 1 & 0.42 & 0.84 \\
\textbf{RF Avg.}&0.50& 0.48& 0.48& 0.09& 0.22& 0.10& 0.22& 0.20& 0.21& 0.11& 0.06& 0.10 & 0.79 & 0.51 & 0.71\\ 			
\textbf{T.A.} & 0.61& 0.41& 0.55& 0.20& 0.26&0.20& 0.39& 0.40& 0.38& 0.25& 0.23& 0.24 & 0.91 & 0.46 & 0.79\\
    \bottomrule
  \end{tabular}
  }
\end{table}

It can be seen that \dfd is the tool that gained the most from considering alternative GTs, with improved RC outcomes in 9 out of the 18 faults where alternative GTs were available. This result is followed by that of  \NL, whose RC results improved for 6 faults, and by GPT-4 with improvements in 4 RC values. In contrast, \DD and \UM only showed RC increases in 2 cases each. To facilitate comparison of tool performance before and after neutrality analysis, Table~\ref{tab:neutrality_results} presents initial average RC, PR, and $F_3$ scores, as well as updated values for both benchmark groups (AF for artificial faults, RF for real faults) and overall (T.A. indicates Total Average). Although \dfd, \NL, and GPT-4 demonstrated notable gains in performance metrics, the comparative rankings of tools based on the original GT align with those observed after neutrality analysis. This is confirmed by the Wilcoxon signed-rank test with $p$-value of 0.002 for the comparison between \dfd and \DD and $p$-value of 0.023 for  \dfd vs \UM. However, the difference between \dfd and \NL does not reach statistical significance, with a $p$-value of 0.066. In general, GPT-4 shows unmatched performance that surpasses the recall of the second-best approach \dfd by 0.32 and 0.29 in artificial and real faults, respectively. %

Our results underscore the value of acknowledging multiple fault causes. FL results change significantly when the ground truth definition is expanded to consider alternative fault-correcting changes.

\begin{tcolorbox}[boxrule=0pt,frame hidden,sharp corners,enhanced,borderline north={1pt}{0pt}{black},borderline south={1pt}{0pt}{black},boxsep=2pt,left=2pt,right=2pt,top=2.5pt,bottom=2pt]
  \textbf{Answer to RQ1.2 (FL Effectiveness After Neutrality Analysis)}: Our neutrality analysis reveals that all FL tools exhibit improved performance when considering alternative ground truths, with \dfd, \NL, and GPT-4 showing the most significant enhancements. This highlights the importance of incorporating alternative ground truths in FL tool evaluation. Furthermore,  results indicate that LLMs provide outstanding assistance in the FL task for DL, outperforming all existing approaches. 
\end{tcolorbox}

\subsection{RQ2 (FL Stability)}
\label{sec:RQ2}

In this RQ, we investigate the stability of the FL results upon experiment repetition. \dfd's output is already calculated from 20 re-trainings to account for instability. \NL does not require any training and is based on static rules that are stable by design. Hence, we did not perform stability analysis on these two tools. We performed 20 runs of all other tools to investigate their stability. We ran \DD and \UM 20 times. We found that their outputs are stable ($\sigma = 0$) across the repetitions of the experiment for \DD and \UM despite the fact that they fully (\DD) or partly (\UM) depend on the dynamics observed in numerous variables inherent to the stochastic model training process.

The situation differs when it comes to GPT-4. Although we minimised the variability in LLM outcomes by setting the temperature parameter of GPT's API to 0, this does not guarantee a deterministic outcome~\cite{ouyang2023llm}. As GPT-4's answers tend to be quite lengthy, we limited the number of repetitions to ten, to make the manual analysis and mapping of the answers to fault types feasible.

When presented with our prompt, GPT-4 generates a numbered list of potential faults (each referred to as an \textit{answer} in the following) affecting the model under test. Across the entire FL benchmark, the average number of answers in GPT-4's outputs is 6.5. For artificial faults, the average number of answers is 6.25, ranging from 5 to 10, and for real faults, it ranges from 4 to 10 with an average of 6.87. The standard deviation of the number of answers across ten repetitions of the same prompt ranges from 0 to 1.99, with an average of 0.98 across the benchmark. It is worth noting that not all answers of GPT can be directly mapped to the fault types we have. Some answers contain assumptions, general recommendations, and best practices without pointing to specific problems. At the same time, in rare cases, it was possible to extract several fault types from a single answer. On average, the extracted fault types constitute 58\% of the answers provided by GPT-4. Interestingly, during the seventh repetition of D7's prompt, GPT-4 has entered an endless loop, repeatedly generating the same set of answers. The output was truncated by the GPT's character limit after 191 answers. As this case is an outlier, we excluded it from our calculations.

On artificial faults,
despite a stable recall equal to 1 (see Table~\ref{tab:neutrality_results_diff}), the average standard deviation $\sigma$ of the $F_3$ metric is 0.04, due to variations in the number of answers across repetitions, impacting precision. 
On real faults, the standard deviation $\sigma$ of the $F_3$ metric is slightly higher at 0.13.
The difference between artificial and real faults might be due to the presence of exactly 1 fault by construction in the former case, while in the latter case the number of faults to be identified ranges from 1 to 5, making the FL task more complex.

\begin{tcolorbox}[boxrule=0pt,frame hidden,sharp corners,enhanced,borderline north={1pt}{0pt}{black},borderline south={1pt}{0pt}{black},boxsep=2pt,left=2pt,right=2pt,top=2.5pt,bottom=2pt]
\textbf{Answer to RQ2 (FL Stability)}: \dfd addresses the randomness of the training process by construction, while the other tools produce stable results across runs. Interestingly, GPT-4 exhibits variable outcomes even with a temperature value of 0, showing higher variability on real-world faults, probably due to their higher complexity.
\end{tcolorbox}

\subsection{RQ3 (FL Efficiency)}
\label{sec:RQ3}

\begin{table}[!h]
  \centering
  \caption{Execution time (in seconds)} 
  \label{tab:fl_time}
   \scalebox{0.8}{
  \begin{tabular}{c|ccccc}
    \toprule
    ID & DFD & DD & NL & UM  &GPT-4\\
    \midrule
    M1 & 605.30 & 6.65 & 7.63 & 37.62   &13.40\\
    M2 & 485.34 & 6.84 & 9.95 & 40.17   &12.75\\
    C1 & 316.10 & 7.34 & 10.02 & 163.08   &15.48\\
    C2 & 338.45 & 7.15 & 9.77 & 4.77   &14.62\\
    C3 & 321.42 & 7.03 & 10.02 & 135.75   &15.57\\
    R1 & 124.50 & 4.75 & 9.44 & 6.25   &16.25\\
    R2 & 115.12 & 4.05 & 9.59 & 5.89   &13.78\\
    R4 & 125.76 & 3.90 & 9.59 & 6.16   &12.51\\
    R5 & 126.13 & 3.58 & 7.7 & 5.10   &10.58\\
    R6 & 133.23 & 4.07 & 9.19 & 6.02   &16.67\\
    R7 & 158.34 & 3.95 & 8.99 & 6.07   &16.58\\
    D1 & 54.50 & 3.30 & 9.85 & 2.07   &12.72\\
    D2 & 451.67 & 20.13 & 9.95 & 18.98   &20.72\\
    D3 & 32.80 & 1.58 & 9.50 & 1.33   &12.91\\
    D4 & 797.46 & 11.66 & 6.87 & 324.57   &18.08\\
    D5 & 562.46 & 11.54 & 7.50 & 27.43   &19.36\\
    D6 & 19.6 & 1.32 & 6.88 & 0.39   &16.67\\
    D8 & 109.40 & 2.36 & 10.16 & 4.38   &15.60\\
    \midrule
    \textbf{Avg.} & \textbf{270.98} & \textbf{6.18} & \textbf{9.03} & \textbf{44.22}   &\textbf{15.24}\\
    \midrule
    M3 & 798.23 & 6.86 & N/A & 38.66   &15.4\\
    R3 & 116.34 & 4.05 & N/A & 6.00   &13.17\\
    D7 & 53.53 & 166.12 & N/A & 1.89   &10.82\\
    D9 & N/A & N/A & 9.35 & 57.07   &12.22\\
    \midrule
\textbf{T.A.} & \textbf{278.37} & \textbf{13.73} & \textbf{9.05} & \textbf{40.89}  &\textbf{14.81}\\
    \bottomrule
  \end{tabular}
  }
\end{table}

In this RQ, we examine the execution time requirements of FL tools, specifically focusing on the time taken for their execution on a given subject. Our measurements exclude preparation and post-processing time, concentrating on the core tool execution.

Table~\ref{tab:fl_time} provides the execution time (in seconds) recorded for a single run of \dfd and \NL, and the average execution time over 20 or 10 runs for \DD and \UM or GPT-4, respectively. The last `T.A.' row outlines the average time each tool takes to perform FL across the whole set of considered faults. For a fair comparison, the `Avg.' row highlights the average time calculated for faults where all tools were applicable. As expected, \dfd requires substantially more time compared to the other tools, as it involves training 20 separate instances for each fault. In contrast, \DD and \UM execute a single retraining iteration, while \NL does not require training a model at all. In addition to dynamic checks, \UM performs a static analysis of the model under test, resulting in a longer execution time compared to \DD. Moreover, \DD often terminates early when a faulty behaviour is detected, leading to the shortest execution times. Interestingly, for faults where training is quick (e.g. C2, D1, D3, and D6), \UM’s complete training phase can be faster than the static analysis performed by \NL. Even if GPT-4's execution time is around 2 times longer than that of the average of the two fastest tools, \DD and \UM, it is still quite low. According to the results of RQ1, GPT-4 delivers the best performance, while being almost 18 times faster than \dfd, which is the second best approach. It is important to note that GPT's API response time depends on a number of factors such as GPT model instance being used, prompt length, API server's  traffic, and client network connection. In our experiments, we noticed very different response times for the same prompt, e.g. for `D2' the response time ranged from 10 to 50 seconds.

\begin{figure}[!h]
    \centering
    \includegraphics[width=0.7\linewidth]{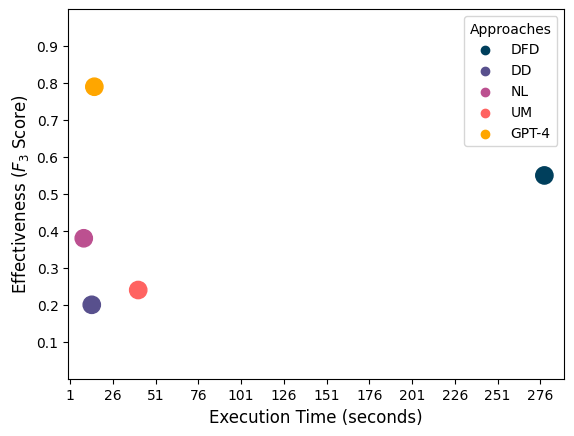}
    \caption{Average execution time and average performance ($F_3$ score) for each tool}
    \label{fig:f3time}
\end{figure}

On average, \DD is the fastest tool, followed by \NL, \UM, and GPT-4, with \dfd being the longest to run. Despite these differences, the execution times for all evaluated tools remain practical for real-world applications. Figure~\ref{fig:f3time} visualises the relationship between each tool’s average execution time and its effectiveness, measured using the $F_3$ score. The visualisation underlines GPT-4's top performance. %

\begin{tcolorbox}[boxrule=0pt,frame hidden,sharp corners,enhanced,borderline north={1pt}{0pt}{black},borderline south={1pt}{0pt}{black},boxsep=2pt,left=2pt,right=2pt,top=2.5pt,bottom=2pt]
\textbf{Answer to RQ3 (FL Efficiency)}: The tools in our study employ different strategies and require a variable number of model re-trainings for fault localisation. GPT-4 delivers the best results requiring a modest execution time of approximately 15 seconds on average. \dfd is the slowest, as it trains 20 instances of the model under test, but it is at the same time the second-best in terms of effectiveness. Notably, none of the tools has a runtime cost that is prohibitively expensive for practical usage.
\end{tcolorbox}

\subsection{RQ4 (Repair Effectiveness)}
\label{sec:RQ4}

\begin{table}[ht]
  \centering
  \caption{Evaluation metric (average: $\mu$; standard deviation: $\sigma$) of faulty model, models patched by Random, \AutoT (AT), HEBO, BOHB, GPTs, and ground truth (GT); regression models are underlined} 
  \label{tab:repair_results}
   \scalebox{0.74}{
  \begin{tabular}{l|r|rr|rr|rr|rr|rr|rr|rr|r}
    \toprule
    Id & Faulty & \multicolumn{2}{c|}{Random} & \multicolumn{2}{c|}{AT} & \multicolumn{2}{c|}{HEBO} & \multicolumn{2}{c|}{BOHB} & \multicolumn{2}{c|}{GPT-3.5} & \multicolumn{2}{c|}{GPT-4} & \multicolumn{2}{c|}{GPT-4T} & \multicolumn{1}{c}{GT} \\
     & Model & $\mu$ & $\sigma$ & $\mu$ & $\sigma$ &  $\mu$ & $\sigma$ &  $\mu$ & $\sigma$ &  $\mu$ & $\sigma$ &  $\mu$ & $\sigma$ &  $\mu$ & $\sigma$ & \\
    \midrule    

    D1 & 0.52  & \textbf{1.00} & 0.00  &  T/O &  T/O & \textbf{0.76} & 0.24  & \textbf{0.95} & 0.14  & \textbf{1.00}  & 0.00 & \textbf{1.00}  & 0.00 & \textbf{1.00}  & 0.00 & \textbf{1.00} \\
    D2 & 0.50  & \textbf{0.67} & 0.00  & \textbf{0.68} & 0.00 & \textbf{0.67} & 0.00  & \textbf{0.67} & 0.01  & \textbf{0.68}  & 0.01 & \textbf{0.73}  & 0.07 & \textbf{0.68}  & 0.02 & \textbf{0.71} \\
    D3 & 0.60  & \textbf{1.00} & 0.01  & \textbf{0.93} & 0.00 & \textbf{1.00} & 0.00  & \textbf{1.00} & 0.00  & \textbf{1.00}  & 0.00 & \textbf{1.00}  & 0.00 & \textbf{1.00}  & 0.00 & \textbf{1.00} \\
    D4 & 0.10  & \textbf{0.95} & 0.02  & 0.10 & 0.00 & \textbf{0.94} & 0.03  & \textbf{0.93} & 0.06  & \textbf{0.98}  & 0.00 & \textbf{0.96}  & 0.02 & \textbf{0.95}  & 0.01 & \textbf{0.94} \\
    D5 & 0.66  & 0.66 & 0.00  &  N/A &  N/A & 0.66 & 0.00  & 0.66 & 0.00  & \textbf{0.69}  & 0.01 & \textbf{0.68}  & 0.01 & \textbf{0.68}  & 0.01 & \textbf{0.75} \\
    D6 & 0.40  & 0.60 & 0.20  &  T/O &  T/O & \textbf{0.85} & 0.21  & \textbf{0.65} & 0.23  & \textbf{0.95}  & 0.15 & \textbf{1.00}  & 0.00 & \textbf{1.00}  & 0.00 & \textbf{1.00} \\
    \underline{D7} & 7.20  & \textbf{0.91} & 1.73  &  N/A &  N/A & \textbf{2.49} & 2.86  & \textbf{0.49} & 1.02  & 4.62 & 3.27 & \textbf{4.16}  & 2.96 & 5.61 & 2.75 & \textbf{0.13} \\
    D8 & 0.28  & \textbf{0.57} & 0.03  & \textbf{0.54} & 0.00 & \textbf{0.57} & 0.02  & \textbf{0.57} & 0.02  & 0.34 & 0.16 & 0.34 & 0.17 & 0.32 & 0.14 & \textbf{0.35} \\
    D9 & 0.10  & \textbf{0.13} & 0.03  & 0.10 & 0.00 & \textbf{0.13} & 0.03  & \textbf{0.12} & 0.01  & 0.10 & 0.00 & 0.10 & 0.00 & 0.10 & 0.00 & \textbf{0.99} \\

    \midrule
C1 & 0.62 & 0.62 & 0.00  & \textbf{0.71} & 0.01 & 0.62 & 0.00  & 0.62 & 0.00  & \textbf{0.68}  & 0.01 & \textbf{0.68}  & 0.01 & \textbf{0.69}  & 0.01 & \textbf{0.70}  \\
C2 & 0.53 & 0.53 & 0.00  &  N/A &  N/A & 0.53 & 0.00  & 0.53 & 0.00  & \textbf{0.68}  & 0.01 & \textbf{0.68}  & 0.01 & \textbf{0.69}  & 0.01 & \textbf{0.70}  \\
C3 & 0.49 & 0.49 & 0.00  &  N/A &  N/A & 0.49 & 0.00  & 0.49 & 0.00  & \textbf{0.68}  & 0.01 & \textbf{0.68}  & 0.01 & \textbf{0.69}  & 0.01 & \textbf{0.70}  \\
\underline{U1} & 0.184 & \textbf{0.152} & 0.051  &  N/A &  N/A & \textbf{0.184} & 0.000  & \textbf{0.184} & 0.000  & \textbf{0.058}  & 0.074 & \textbf{0.032}  & 0.001 & \textbf{0.032}  & 0.001 & \textbf{0.046}  \\
\underline{U2} & 0.124 & \textbf{0.052} & 0.059  &  N/A &  N/A & \textbf{0.004} & 0.000  & 0.064 & 0.060  & 0.058 & 0.074 & \textbf{0.032}  & 0.001 & \textbf{0.032}  & 0.001 & \textbf{0.046}  \\
\underline{U3} & 0.150 & 0.069 & 0.069  &  N/A &  N/A & \textbf{0.034} & 0.058  & 0.101 & 0.065  & \textbf{0.058}  & 0.074 & \textbf{0.032}  & 0.001 & \textbf{0.032}  & 0.001 & \textbf{0.046}  \\
\underline{U4} & 0.398 & 0.087 & 0.126  &  N/A &  N/A & \textbf{0.004} & 0.000  & \textbf{0.004} & 0.000  & \textbf{0.058}  & 0.074 & \textbf{0.032}  & 0.001 & \textbf{0.032}  & 0.001 & \textbf{0.046}  \\
\underline{U5} & 0.071 & 0.071 & 0.000  &  N/A &  N/A & 0.071 & 0.000  & 0.071 & 0.000  & 0.058 & 0.074 & \textbf{0.032}  & 0.001 & \textbf{0.032}  & 0.001 & \textbf{0.046}  \\
\underline{U6} & 0.134 & 0.082 & 0.063  &  N/A &  N/A & 0.082 & 0.063  & \textbf{0.043} & 0.059  & 0.058 & 0.074 & \textbf{0.032}  & 0.001 & \textbf{0.032}  & 0.001 & \textbf{0.046}  \\
\underline{U7} & 0.096 & \textbf{0.023} & 0.037  &  N/A &  N/A & \textbf{0.032} & 0.042  & \textbf{0.032} & 0.042  & 0.058 & 0.074 & \textbf{0.032}  & 0.001 & \textbf{0.032}  & 0.001 & \textbf{0.046}  \\
\underline{U8} & 0.163 & 0.163 & 0.000  &  N/A &  N/A & 0.163 & 0.000  & 0.163 & 0.000  & \textbf{0.058}  & 0.074 & \textbf{0.032}  & 0.001 & \textbf{0.032}  & 0.001 & \textbf{0.046}  \\
M1 & 0.85 & 0.86 & 0.04  &  N/A &  N/A & \textbf{0.94} & 0.04  & 0.87 & 0.04  & \textbf{0.99}  & 0.00 & \textbf{0.99}  & 0.00 & \textbf{0.99}  & 0.00 & \textbf{0.99}  \\
M2 & 0.11 & \textbf{0.43} & 0.36  & \textbf{0.99} & 0.00 & \textbf{0.93} & 0.04  & 0.21 & 0.26  & \textbf{0.99}  & 0.00 & \textbf{0.99}  & 0.00 & \textbf{0.99}  & 0.00 & \textbf{0.99}  \\
M3 & 0.10 & \textbf{0.52} & 0.41  & \textbf{0.32} & 0.03 & \textbf{0.87} & 0.25  & \textbf{0.45} & 0.35  & \textbf{0.99}  & 0.00 & \textbf{0.99}  & 0.00 & \textbf{0.99}  & 0.00 & \textbf{0.99}  \\
R1 & 0.51 & 0.56 & 0.10  & \textbf{0.58} & 0.00 & 0.52 & 0.03  & 0.52 & 0.02  & \textbf{0.81}  & 0.01 & \textbf{0.81}  & 0.01 & \textbf{0.81}  & 0.01 & \textbf{0.82}  \\
R2 & 0.39 & \textbf{0.67} & 0.09  & 0.23 & 0.00 & \textbf{0.50} & 0.12  & \textbf{0.71} & 0.10  & \textbf{0.81}  & 0.01 & \textbf{0.81}  & 0.01 & \textbf{0.81}  & 0.01 & \textbf{0.82}  \\
R3 & 0.37 & \textbf{0.68} & 0.10  & 0.34 & 0.00 & 0.53 & 0.19  & \textbf{0.64} & 0.14  & \textbf{0.81}  & 0.01 & \textbf{0.81}  & 0.01 & \textbf{0.81}  & 0.01 & \textbf{0.82}  \\
R4 & 0.67 & 0.70 & 0.05  & \textbf{0.81} & 0.00 & 0.67 & 0.03  & \textbf{0.73} & 0.06  & \textbf{0.81}  & 0.01 & \textbf{0.81}  & 0.01 & \textbf{0.81}  & 0.01 & \textbf{0.82}  \\
R5 & 0.63 & \textbf{0.72} & 0.06  & \textbf{0.82} & 0.00 & \textbf{0.69} & 0.07  & \textbf{0.71} & 0.05  & \textbf{0.81}  & 0.01 & \textbf{0.81}  & 0.01 & \textbf{0.81}  & 0.01 & \textbf{0.82}  \\
R6 & 0.52 & \textbf{0.75} & 0.04  & \textbf{0.56} & 0.00 & 0.62 & 0.11  & \textbf{0.72} & 0.08  & \textbf{0.81}  & 0.01 & \textbf{0.81}  & 0.01 & \textbf{0.81}  & 0.01 & \textbf{0.82}  \\
R7 & 0.28 & \textbf{0.68} & 0.13  & 0.12 & 0.00 & \textbf{0.61} & 0.12  & \textbf{0.67} & 0.09  & \textbf{0.81}  & 0.01 & \textbf{0.81}  & 0.01 & \textbf{0.81}  & 0.01 & \textbf{0.82}  \\

    \bottomrule
  \end{tabular}
  }
\end{table}

\begin{figure}[h]
  \centering
  \includegraphics[width=\linewidth]{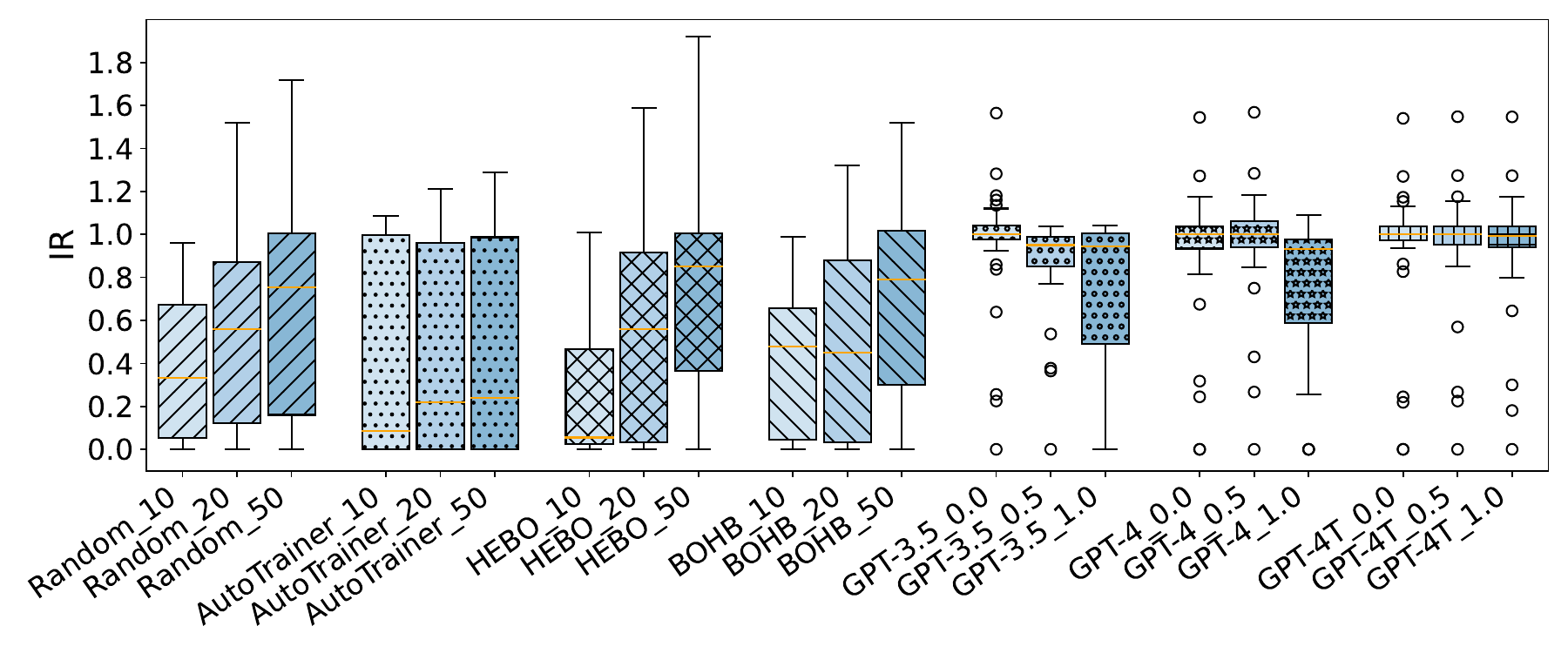}
  \caption{IR values from all faults in the benchmark, broken down by the
   combinations of repair technique and budget, shown as [technique]\_[budget] for Random, \AutoT, and HPO and [version]\_[temperature] for GPTs. Note that some IR values are higher than 1.0, meaning that the corresponding patches are better than the ground truth patches.}
  \label{fig:IR}
\end{figure}

Table~\ref{tab:repair_results} shows the evaluation metric value (accuracy or regression loss, depending on the model; regression models are underlined) of the patched models averaged over ten runs of patch generation ($\mu$) for Random, \AutoT (AT), HEBO, BOHB, and GPTs. Column `Faulty Model' shows the metric value for the initial faulty model, while column `GT' shows the value for the ground truth repaired model. The cases that exhibit the statistical significance of the difference between the metric value of the faulty model and patched model are highlighted in \textbf{bold}. `N/A' means that \AutoT cannot be applied to the faulty program (e.g., to UnityEyes, which is a regression model) or did not find any failure symptoms, and `T/O' means that \AutoT did not have enough time to find any patch. Note that Table~\ref{tab:repair_results} only shows the results for the time budget 20 for Random and HPO and temperature 0.5 for GPTs. For the full tables, please consult the online supplementary material at \url{https://github.com/testingautomated-usi/dl-fl-repair}.

Overall, all three GPTs exhibit competitive performance relative to ground truth patches. Out of all 30 subject faults except D8 and D9, GPTs find patches that are statistically better than the faulty models. These trends persist across different temperature values, demonstrating the robustness of their performance. Both HEBO and BOHB can find patches in 19 cases, followed by Random with 17 cases, and \AutoT with 10 cases.

Figure~\ref{fig:IR} shows IR values for the considered techniques: within the 20 trainings time budget for Random and HPO and with 0.5 temperature for GPTs, the median IR values of both GPT-4 and GPT-4T are 1.0, followed by GPT-3.5 with 0.95, both Random and HEBO with 0.55, BOHB with 0.45, and \AutoT with 0.18. This indicates that, in general, \AutoT and HPO techniques fail to generate more effective patches than Random. Despite being a baseline technique, overall, Random performs surprisingly well. This conclusion differs from the ones reported in the papers of HEBO and BOHB~\cite{Cowen-Rivers2022lm, bohb}, which showed that their techniques are better than Random. We hypothesise that this is due to the different set of subjects that we considered, which has a larger number of hyperparameters and correspondingly a larger search space: our study required tuning  an average of 15 hyperparameters as opposed to the six of their studies. Across all temperatures and versions, GPTs consistently exhibit superior performance with lower variance in terms of IR values. While higher temperatures tend to increase variance, the latest version, GPT-4T, shows consistently low variance even at a high temperature (i.e., 1.0). This result suggests that GPTs are adept at recommending generally good repair suggestions for a given task, dataset, and model structure. Such ability remains a key advantage of GPTs, regardless of the search space size of a given problem, while competing techniques suffer when the search space grows.

\begin{tcolorbox}[boxrule=0pt,frame hidden,sharp corners,enhanced,borderline north={1pt}{0pt}{black},borderline south={1pt}{0pt}{black},boxsep=2pt,left=2pt,right=2pt,top=2.5pt,bottom=2pt]
\textbf{Answer to RQ4 (Repair Effectiveness)}:
The random baseline produces comparable or better patches than HPO and \AutoT, but the effectiveness of tools varies depending on the fault. Generally, GPTs exhibit stable and superior performance, often producing patches that are competitive with the ground truth ones.
\end{tcolorbox}

\subsection{RQ5 (Repair Stability)}
\label{sec:RQ5}

Table~\ref{tab:repair_results} shows also the standard deviations ($\sigma$) of the evaluation metric, used to quantify the stability of the patches found by each tool across ten runs (i.e., $\sigma$ quantifies the performance variability of the best patched model across multiple executions of each tool). Below, we comment on the standard deviation of each tool, considering only the cases showing statistical significance of the model performance improvement. 

Among all the techniques, GPTs demonstrate the smallest average standard deviation: 0.012 (GPT3.5), 0.015 (GPT-4T), and 0.023 (GPT-4). This is followed by \AutoT with 0.034, HEBO at 0.327, BOHB at 0.409, and Random with the highest deviation of 0.474, which is quite expected. These results indicate that GPTs exhibit remarkable stability in their recommendations, even considering the inherent randomness introduced by temperature-based diversity. Thus, GPTs offer a valuable advantage to developers in terms of both repair effectiveness and stability across multiple runs.
\AutoT also exhibits a stable effectiveness since the number of repair operators being applied is relatively small compared to the others (see their coverages in Tables~\ref{tab:AFmodels} \& \ref{tab:RFmodels} and complexities in Section~\ref{sec:RQ7} for details), allowing it to generate consistent patches across executions. In contrast, HPO techniques, as well as Random, tend to produce more diverse and different patches, which implies that their patches are less stable in terms of patched model performance. %

\begin{tcolorbox}[boxrule=0pt,frame hidden,sharp corners,enhanced,borderline north={1pt}{0pt}{black},borderline south={1pt}{0pt}{black},boxsep=2pt,left=2pt,right=2pt,top=2.5pt,bottom=2pt]
\textbf{Answer to RQ5 (Repair Stability)}:
GPTs are shown to be the most stable technique, consistently generating stable patches across multiple runs. \AutoT also exhibits a relatively low standard deviation due to its selective application of operators from a limited set. In contrast, HPO techniques and Random produce more varied patches, making them susceptible to instability.
\end{tcolorbox}

\subsection{RQ6 (Repair Efficiency)}
\label{sec:RQ6}

Automated program repair for traditional software usually requires a significant amount of time and computational resources, as it needs to search a large space of patches while running the tests for each candidate patch. Techniques such as Random and HPO also have similar issues because each patch requires training and validating the model from scratch. In contrast, GPTs employed in our study offer a straightforward approach to patch generation without any iterative search and refinement process, which hence does not depend on the search execution budget. Correspondingly, this RQ focuses only on the impact of varying budgets on the performance of search-based techniques, i.e., Random and HPO, to provide insights into their efficiency.

We investigate three different time limits, 10, 20, and 50, under the
assumption that developers may have different time constraints when repairing a faulty DL model. We report only a time limit of 20 in Table~\ref{tab:repair_results} (full results are available in the online supplementary material at \url{https://github.com/testingautomated-usi/dl-fl-repair}). As expected, all techniques produce more patches showing a statistical significance of the improvements when larger budgets are allowed. For instance, Random finds patches showing statistical significance in 14 cases with a 10 time budget, which becomes 17 cases with a 20 time budget and 23 cases with a 50 time budget. This trend is consistent even considering IR, as shown in Figure~\ref{fig:IR}: larger time budget results in larger IR as well as a smaller standard deviation. \AutoT does not take advantage so much of a larger time budget, compared to the other techniques, due to its limited search space. HEBO can be a good alternative to Random when the budget is as large such as 50: it shows slightly better performance than Random with a smaller standard deviation.

\begin{tcolorbox}[boxrule=0pt,frame hidden,sharp corners,enhanced,borderline north={1pt}{0pt}{black},borderline south={1pt}{0pt}{black},boxsep=2pt,left=2pt,right=2pt,top=2.5pt,bottom=2pt]
\textbf{Answer to RQ6 (Repair Efficiency)}:
Using a larger time budget results in more stable and better patches. The results also show that \AutoT does not benefit from larger budgets, while HPO techniques can benefit from them.
\end{tcolorbox}

\begin{figure}[!h]
  \begin{subfigure}{\linewidth}
      \centering
      \includegraphics[width=\linewidth]{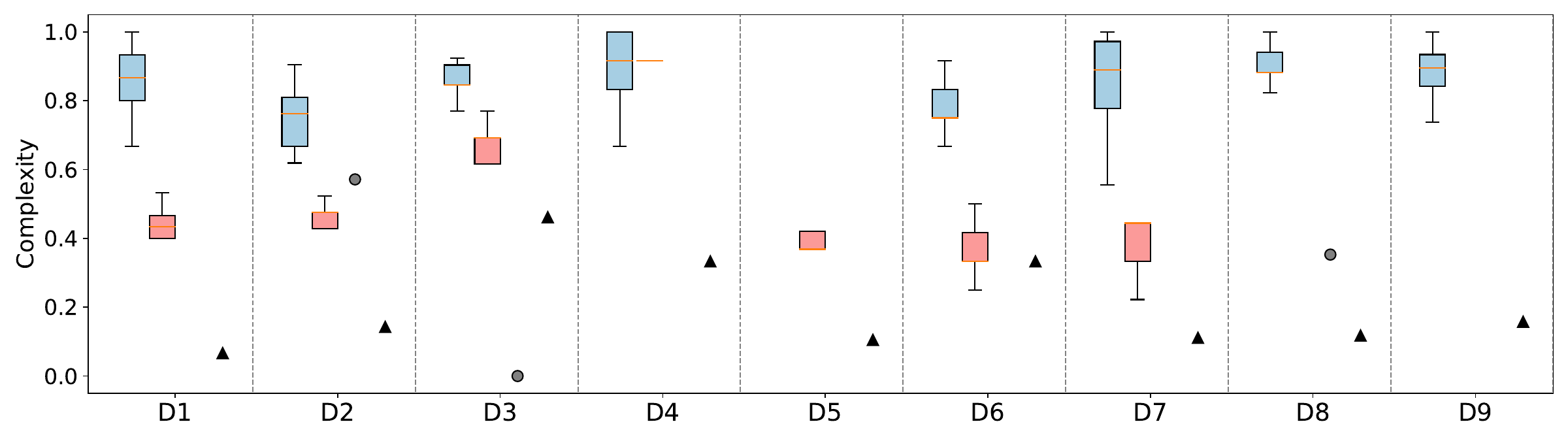}
      \caption{Real faults}
      \label{fig:complexity_plot1}
  \end{subfigure}
  \begin{subfigure}{\linewidth}
      \centering
      \includegraphics[width=\linewidth]{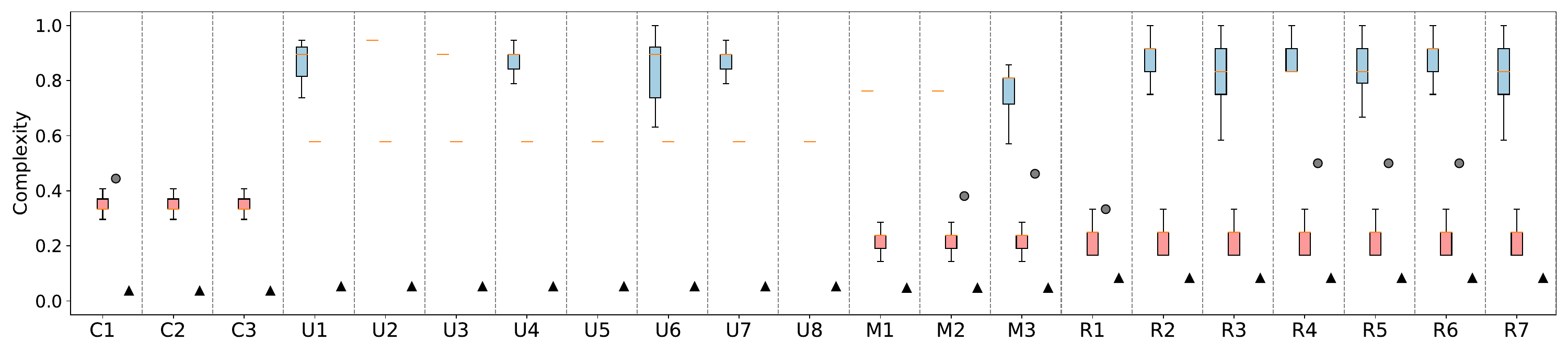}
      \caption{Artificial faults}
      \label{fig:complexity_plot2}
  \end{subfigure}
  \caption{Complexity of statistically significant patches. The \colorbox{boxcolorblue}{blue} boxplots represent HPO and Random's patches, \colorbox{boxcolorred}{red} boxplots GPTs' ones, triangles the ground truth's ones, and circles the \AutoT's ones.}
  \label{fig:complexity_plot}
\end{figure}

\subsection{RQ7 (Patch Complexity)}
\label{sec:RQ7}

Figure~\ref{fig:complexity_plot} presents the boxplots of the complexity of the statistically significant patches generated by HPO techniques and Random with time budget 20 (shown in blue boxplots) and GPTs (shown in red boxplots). The triangles and circles show the complexity of the ground truth patches and \AutoT's patches, respectively.\footnote{We present integrated results of Random and two HPO techniques as they all show similar trends, and we do not use boxplots for ground truth and \AutoT as their variance is too small. Also, note that there are missing boxplots and circles because we only consider statistically significant patches.\label{fn:complexity}}
Overall, the complexity of the generated patches of HPO and Random is much higher than the complexity of the ground truth patches. This means that the generated patches manipulate many different hyperparameters (around 80\% to 90\% of them) to achieve a significant improvement of the faulty model. The ground truth patches make fewer changes, despite achieving similar or higher evaluation metric values. The main reason for this difference is that both HPO and Random explore the hyperparameter space at large in search for configurations that improve the model's accuracy. Random is completely unconstrained in its exploration: thus, it is expected that it can generate solutions that are far from the initial faulty model. HPO, on the other hand, balances exploitation (i.e., local improvements of the best model found so far, which at the beginning is the initial faulty model) and exploration (i.e.,  sampling of new diversified points in the hyperparameter space to avoid getting stuck in a local minimum). Consequently,  results suggest that, in our subjects, the exploration component of search-based approaches is dominant, and improvements are obtained only when HPO techniques move away from the initial model. 

The complexity of the patches generated by GPTs is lower than those produced by Random and HPO techniques, although they remain more complex than the ground truth patches. While GPTs can generate competitive patches in terms of model performance, the results on their complexity indicate that there is room for simplifying the patches:, which would enhance the developer's understanding and acceptance of the recommendations made by the GPTs.

The patches generated by \AutoT have lower complexity than Random and HPO. This is consistent with its design principle: it can handle a narrow set of repair actions, targeting specific fault types, which makes the tool either effective and capable of improving the initial solution with a small number of changes or completely ineffective.

\begin{figure}[!h]
  \begin{subfigure}{\linewidth}
      \centering
      \includegraphics[width=\linewidth]{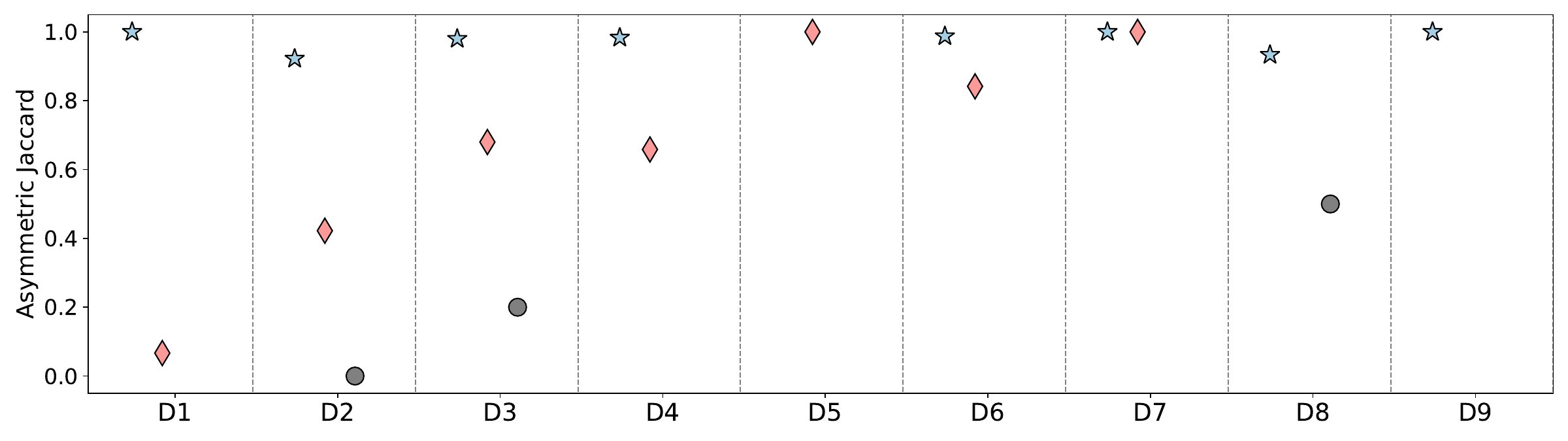}
      \caption{Real faults}
      \label{fig:aj_plot_deepfd}
  \end{subfigure}
  \begin{subfigure}{\linewidth}
      \centering
      \includegraphics[width=\linewidth]{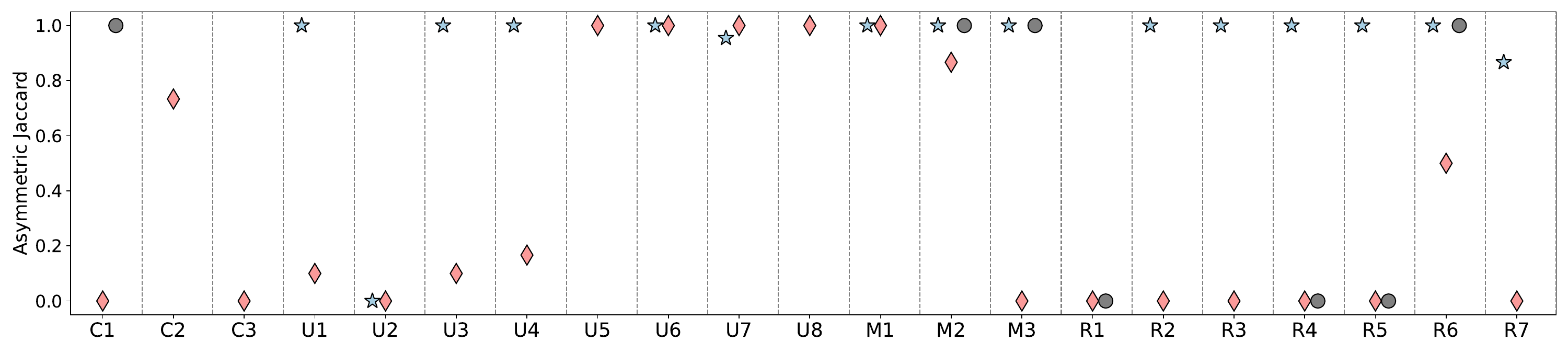}
      \caption{Artificial faults}
      \label{fig:aj_plot_deepcrime}
  \end{subfigure}
  \caption{Asymmetric Jaccard of statistically significant patches. The \colorbox{boxcolorblue}{blue} stars represent HPO and Random's patches, \colorbox{boxcolorred}{red} diamonds GPTs' ones, and circles the \AutoT's ones.}
  \label{fig:aj_plot}
\end{figure}

Figure~\ref{fig:aj_plot} shows AJ values of each technique: Despite the generally higher complexity, the observed AJ values suggest that the generated patches do contain the same \emph{ingredients} as the ground truth patches,  i.e., they include similar repair operators. Specifically, the AJ values for HPO and Random attain a mean of 0.97 for real faults and 0.85 for artificial faults, respectively. This suggests that these patches may be \emph{bloated}, incorporating redundant changes. Conversely, GPTs show lower AJ values, with 0.62 for real faults, and a notably lower 0.18 for artificial faults. This divergence suggests that GPTs tend to generate diverse patches, 
which truly represent alternative solutions to the problem.
In contrast, \AutoT displays a narrower repair scope, reflected in its AJ values of 0.26 for real faults and 0.57 for artificial faults, respectively. However, when considering alternative ground truths, the AJ values for both GPTs and \AutoT increase notably, with GPTs achieving a mean of 0.95 (+0.33) for real faults and 0.92 (+0.78) for artificial faults, while \AutoT reaches 0.96 (+0.70) and 0.51 (+0.06), respectively. This finding underscores the importance of accounting for alternative ground truths, even in the evaluation of DL repair techniques.

\begin{tcolorbox}[boxrule=0pt,frame hidden,sharp corners,enhanced,borderline north={1pt}{0pt}{black},borderline south={1pt}{0pt}{black},boxsep=2pt,left=2pt,right=2pt,top=2.5pt,bottom=2pt]
\textbf{Answer to RQ7 (Patch Complexity)}:
The complexity of the patches generated by HPO techniques and Random is high compared to that of the GPT and \AutoT patches, and compared to the ground truth. All techniques tend to reuse a large proportion of ingredients from the ground truth patch, possibly bloated with other redundant, irrelevant changes.
\end{tcolorbox}

%% file: discussion.tex
\section{Discussion}
\label{sec:discussion}

\subsection{Why do LLMs excel in DL repair and FL?}

Our empirical study showd that a family of GPT models outperformed state-of-the-art DL repair and FL techniques from SE and ML. Interestingly, we found that they excel at these tasks with quite basic prompts, without any need for sophisticated prompting techniques like few-shot learning or Retrieval-augmented generation (RAG). We attribute this to LLMs' outstanding ability to predict repetitive patterns, which aligns well with localising and fixing common DL faults related to the DL model architecture. In contrast, FL and repair in \textit{traditional software} may pose greater challenges for LLMs, as they require understanding logic, semantics, and possible executions of the given program. However, our scope on DL faults primarily involves parameter modifications at a syntactic level, which we believe is the key reason for LLMs' superior performance. 
Moreover, given the paper's objective of evaluating FL and repair techniques, we initially intended to explore the potential synergy between these processes by feeding the output of FL techniques into repair techniques. However, since LLMs demonstrated near-perfect performance in repairing our subject DL programs without FL assistance, we concluded that investigating this potential bridge would not yield significant insights.

\subsection{Benchmark}
The analysis of the existing benchmarks of real faults currently used in the literature has revealed that the majority of real faults collected so far are rather simplistic. In many cases,  benchmark models represent toy examples for naive tasks. The datasets used to train and test them are often either randomly generated or small. For this reason, we deemed the addition of the artificial faults produced by \dc as quite important and useful, since they cover a larger variety of fault types and affect more diverse and complex models. A common pattern we observed in the results is that on large models such as CIFAR10 model, all techniques except for GPT could not generate any successful fixes (see C2 and C3 in Table~\ref{tab:repair_results}). Since the number of hyperparameters of the CIFAR10 model is 27, which is twice bigger than that of the Reuters model, the search space is relatively large, so it becomes more difficult to find patches.

\subsection{Patch Minimisation}
Compared to traditional APR techniques for source code, one critical step that is missing in DL repair is \textit{patch minimisation}. Although our analysis shows that smaller patches do exist and such patches are useful for developers, minimization might be computationally expensive due to the stochastic nature of model training. Patch minimisation for DL faults remains an unexplored area.

%% file: related_work.tex
\section{Related Work}
\label{sec:related_work}

\subsection{Automated DL Fault Localisation}

Fault localisation in DL models is an emerging field within DL testing~\cite{deeplocalize, wardat2022deepdiagnosis, Cao2022zz, nikanjam2021automatic, schoop2021umlaut}. The majority of proposed approaches have focused on analysing the run-time behaviour during model training. Based on collected information and predefined rules, these methods determined and reported abnormalities~\cite{deeplocalize, wardat2022deepdiagnosis, schoop2021umlaut}.

During model training, \DD~\cite{wardat2022deepdiagnosis}, which was built on top of \DL~\cite{deeplocalize}, utilises a callback to gather performance indicators such as loss function values, weights, gradients, and activations. These tools then compare the analysed values with predefined failure symptoms. \UM~\cite{schoop2021umlaut} combines static checks of model structure and parameters with dynamic monitoring of training and model behaviour. It enhances the check results with error message analysis, providing best practices and suggestions on how to deal with the faults.

Unlike previously discussed methods, \NL~\cite{nikanjam2021automatic} is a model-based approach that employs meta-modelling and graph transformations for fault detection. Given a model under test, it constructs a meta-model consisting of a base skeleton and some fundamental properties. This model is then checked against a set of 23 rules embodied in graph transformations, each representing a fault or a design issue. \dfd~\cite{Cao2022zz} employs mutation testing to construct a database of mutants and original models, to train a fault type ML classifier. From the mutants, it extracts a number of runtime features and uses several combinations of them to localise the faults. Our empirical evaluation of FL techniques includes \DD, \UM, \dfd and \NL. Results show that they are outperformed by LLM-based fault localisation.

\subsection{Automated DL Repair}

Several DL repair approaches come from ML, where they belong to the hyperparameter optimization family. A notable exception from software engineering is \AutoT, a tool that continues to train an already trained model using patched hyperparameters. The goal of our empirical study for repair tools was to compare these two families of approaches with LLM-based repair for DL. No previous empirical study attempted to conduct any similar comparison.

On the other hand, post-training, model-level repair of DL, i.e., repair through the modification of the weights of an already trained model, is gaining increasing popularity. \textit{CARE}~\cite{sun2022causality} identifies and modifies weights of neurons that contribute to detected model misbehaviour until the defects are eliminated. \textit{Arachne}~\cite{Sohn2022cr} operates similarly to \textit{CARE} while ensuring the non-disturbance of the correct behaviour of a model under repair. \textit{GenMuNN}~\cite{wu2022genmunn} ranks the weights based on the effect on predictions. Using the computed ranks, it generates mutants and evaluates and evolves them using a genetic algorithm. \textit{NeuRecover}~\cite{tokui2022neurecover} keeps track of the training history to find the weights that have changed significantly over time. Such weights become subject for repair if they are not beneficial for the prediction of the successfully learnt inputs but have become detrimental for the inputs that were correctly classified in the earlier stages of the training. Similarly to NeuRecover, \textit{I-Repair}~\cite{henriksen2022repairing} focuses on modifying localised weights to influence the predictions for a certain set of misbehaving inputs, whereas minimising the effect on the data that was already correctly classified. \textit{NNrepair}~\cite{usman2021nn} adopts fault localisation to pinpoint suspicious weights and treats them by using constraint solving, resulting in minor modifications of weights. 

\textit{PRDNN}~\cite{sotoudeh2021provable} took a slightly different path by focusing on the smallest achievable single-layer repair. If provided with a limited set of problematic inputs and a model, this algorithm returns a repaired DNN that produces correct output for these and similar inputs and retains the model's behaviour for other, dissimilar kinds of data. \textit{Apricot}~\cite{Zhang2019zj} uses a DL model trained on a reduced subset of inputs and then uses the weights of the reduced model to adjust the weights of the full model to fix its misbehaviour on the inputs from the reduced dataset. In our work, we are interested in the approaches that recommend changes to the model's source code rather than patching the weights of the model.

\subsection{LLMs and Their Usage}

LLMs have been adopted for a diverse range of software engineering tasks~\cite{fan2023large}, including code generation~\cite{jiang2024survey, liu2024your, du2024evaluating, ni2024l2ceval}, testing~\cite{wang2024software,liu2024llm,yu2023llm}, fault localisation~\cite{wu2023large,liu2024empirical} and program repair~\cite{zhang2024systematic, jin2023inferfix, cao2023study, xia2023automated}. Employing LLMs effectively often entails fine-tuning with additional data or iterative prompt engineering to improve the output. Moreover, their stochastic nature should be accounted for when experimenting with LLMs~\cite{ouyang2023llm}. 

Most existing works on applying LLMs to software engineering tasks focus on traditional software and not DL systems. The study by Cao et al.~\cite{cao2023study} is the work most relevant to ours. The paper focused on debugging DL programs. Debugging is regarded as a complex activity consisting of three subtasks: fault detection, fault localisation, and repair. The evaluation is performed for each of the three subtasks and on 34 programs from the \dfd dataset~\cite{deepfd}. The authors compare the performance of LLMs with baseline tools like \AutoT and \dfd. This study also explores how prompt design and the use of LLM dialogue mode impact LLM performance.

Our work shares some similarities with Cao et al.~\cite{cao2023study}, as both focus on applying LLMs to similar tasks for DL programs. However, there are many key differences, which show how are study represents a substantial advancement of the state of the art: (1) Cao et al.~\cite{cao2023study} exclusively used the \dfd dataset without applying any fault exclusion criteria. In contrast, we meticulously analysed the \dfd dataset and implemented filtering steps (as discussed in Section~\ref{sec:real_faults}) to include only verified and reproducible faults in our study. (2) We generated DL programs with artificially injected faults using nine mutation operators across six DL systems, ensuring higher diversity in program domains and fault types. (3) Cao et al.~\cite{cao2023study} based their analysis on a single GT, whereas we generated multiple alternative GTs for each fault and reported results across these alternatives to account for various ways to improve the DL system. (4) Our study encompassed all relevant state-of-the-art tools and compared LLMs with four FL and three repair tools, while they focused on only two baseline tools (\AutoT and \dfd).

%% file: threats_to_validity.tex
\section {Threats to Validity}
\label{sec:threats}

\subsection{Construct Threats}
Threats to construct validity are due to the measurement of the effectiveness of the FL tools and the interpretation of the tools' output. We use a simple count of the matches between FL results and the ground truth, along with the RC, PR and $F_{\beta}$ metrics that are standard in information retrieval. For repair tools, accurately measuring their performance can be challenging: all evaluation metrics used in the benchmark are standard and widely adopted in the literature.

\subsection{Internal Threats}

Threats to internal validity include the selection of evaluated approaches. We considered all state-of-the-art techniques and their publicly available implementations to the best of our knowledge. For repair tools, the selection of HPO algorithms is crucial. We studied the state-of-the-art HPO algorithms, selecting novel and top-performing methods along with established baselines. To ensure correct implementations, we relied on widely used libraries and frameworks. LLMs face the challenge of data leakage, where the DL programs in our benchmark might exist in the training data. However, we partly addressed this by assessing the tools using artificially seeded faults, guaranteeing their absence from the training set.

\subsection{External Threats}

To address external validity, we meticulously selected faults from both artificial and real sources, encompassing a diverse range of subjects for evaluating FL and repair techniques. All faults in our benchmark were obtained through a rigorous selection process.

%% file: conclusion.tex
\section{Conclusion}
\label{sec:conclusion}

In this paper, we present a thorough evaluation of state-of-the-art fault localisation and repair techniques for deep learning models, revealing their strengths and weaknesses. We introduced a novel approach utilising large language models for both fault localisation and repair tasks, which outperformed existing techniques. This underscores the potential of LLMs as a promising research direction and offers practical solutions for FL and repair in DL models. Furthermore, our curated benchmark provides valuable insights into the current landscape of FL and repair techniques, emphasising the need for a more comprehensive evaluation that considers multiple ground truth patches.

%% file: appendix.tex
\appendix
\section{RQ1 Table For Each Tool} \label{sec:RQ1_appendix}

Tables~\ref{tab:main_results_fl_1_4_1}, \ref{tab:main_results_fl_1_4_2}, \ref{tab:main_results_fl_1_4_3}, \ref{tab:main_results_fl_1_4_4}, and \ref{tab:main_results_fl_openai} report the outputs of FL tools (\dfd, \DD, \NL, \UM, and GPT-4, respectively) when applied to our set of benchmark faults. Column `Id' indicates fault identification code,  while column `GT' (i.e., `Ground Truth') lists faults affecting the buggy version of each fault and column `\#F' reports the number of such faults. For example, C1 is affected by one fault 'ACH(2)' that stands for the sub-optimal selection of the activation function for the third layer (which has index = 2) of the neural network. Column `Matches-GT', in its turn, for each fault of the ground truth, shows whether it was detected by a FL tool or not (1 if yes and 0 otherwise). Correspondingly, column `\#M' counts the number of detected faults by the tool. For each row (fault), this number is underlined if it is the best result achieved across all the compared approaches. For each tool and for each fault source (artificial injection or real-world) we provide the average number of GT faults and the average number of detected faults (rows `Avg.' for each fault source; row `T.A.', i.e., Total Average, for the overall benchmark).  

In ground truth, for faults affecting layers, such as selection of activation function, we provide the indexes of all faulty layers in round brackets after the fault type abbreviation (i.e. ACH(2)). For tools that can pinpoint faults to specific layers, the same information is specified in the same manner in the `<tool\_name>-output' column that contains the localised fault list generated. The cases when a FL tool was not able to identify any faults in the model under test are marked by `-'. We use `N/A' to specify that a tool was not applicable to model under test or crashed during the execution. For example, \NL accepts only optimisers that are defined as strings (e.g., `adam'), which automatically implies that the framework will use the default learning rate for the selected optimiser. It would be impossible for \NL to find an optimiser with a custom learning rate. %
Typically, we use comma (`,') to separate all detected faults. In some cases,  a vertical bar (`|') is used to illustrate that a tool has suggested several alternative fault types, i.e. the tool suggests either of them could be the possible cause of model's misbehaviour. 

Notably, in most instances, \UM (20 out of 22) and \DD (15 out of 22) recommend modifying the last layer's activation function to `softmax', despite this function already being `softmax' in 73\% of the \UM cases and 67\% of the \DD cases. A similar recommendation occurs once with \NL.  We filter out these misleading suggestions from the tools' output. Additionally, \UM occasionally warns of potential overfitting. Since this is a precautionary message rather than a direct indication of a specific fault, we also omit it from our analysis. The complete output from the tools can be found in our replication package.

\begin{table*}[htb]
  \centering
  \caption{Ground Truth (GT) and FL outcome generated by \textbf{\dfd (DFD)}; \#F indicates the number of ground truth faults, while \#M the number of ground truth faults detected by the tool (with underline used to indicate the best result among all tools being compared). Avg. shows the average within artificial or real faults. T.A. shows the total average across faults.} 
  \label{tab:main_results_fl_1_4_1}
   \scalebox{0.82}{
  \begin{tabular}{l|l|c|l|c|l}
    \toprule
    Id & GT  &\#F& Matches-GT  &\#M& DFD-output \\
    \midrule    
M1 & WCI(0)  &1& 0  &0 & HLR, ACH, LCH, HNE  \\
M2 & ACH(7)  &1& 0  &0 & OCH, HLR, HNE  \\
M3 & HLR  &1& 1  &\underline{1} & OCH, HLR, LCH  \\
C1 & ACH(2)  &1& 1  &\underline{1} & OCH, HLR, ACH, LCH  \\
C2 & HNE  &1& 0  &0 & OCH, ACH, LCH  \\
C3 & WCI(2)  &1& 0  &0 & OCH, ACH, LCH, HNE  \\
R1 & RAW(0)  &1& 0  &0 & HLR, LCH, HNE  \\
R2 & ACH(2)  &1& 0  &0 & OCH, LCH, HNE  \\
R3 & HLR  &1& 0  &0 & OCH, LCH, HNE  \\
R4 & LCH  &1& 1  &\underline{1} & ACH, LCH  \\
R5 & OCH  &1& 1  &\underline{1} & OCH, ACH, HNE  \\
R6 & WCI(0)  &1& 0  &0 & OCH, ACH, LCH, HNE  \\
R7 & ACH(2)  &1& 0  &0 & OCH, LCH, HNE  \\
\textbf{Avg.}& & \textbf{1}& & \textbf{0.3}& \\
    \midrule
D1 & ACH(7)  &1& 1  &\underline{1} & ACH  \\
D2 & OCH, HNE,  HBS  &3& 0, 0, 0  & 0 & ACH  \\
D3 & OCH, LCH, ACH(0,1), HNE, HBS  & 5 & 1, 0, 0, 0, 0  &1 & OCH, HLR  \\
D4 & ACH(0,1), LCH, HLR  & 3 & 0, 0, 0  & 0 & OCH  \\
D5 & HNE, HBS  & 2 & 0, 0  & 0 & OCH, ACH  \\
D6 & HLR, HNE, LCH, ACH(1)  &4& 1, 1, 0, 0  &\underline{2} & OCH, HLR , HNE  \\
D7 & HLR  &1& 0  &0 & LCH  \\
D8 & OCH, HLR  & 2 & 1, 1  &\underline{2} & OCH, HLR, LCH, HNE  \\
D9 & CPP, ACH(5,6), HBS  &3& 0, 0, 0  & 0 & N/A  \\
 \textbf{Avg.}& & \textbf{2.7}& & \textbf{0.7}&\\
     \midrule
 \textbf{T.A.}& & \textbf{1.7}& & \textbf{0.5}&\\
\bottomrule
  \end{tabular}
  }
\end{table*}

\begin{table*}[htb]
  \centering
  \caption{Ground Truth (GT) and FL outcome generated by \textbf{\DD (DD)}; \#F indicates the number of ground truth faults, while \#M the number of ground truth faults detected by the tool (with underline used to indicate the best result among all tools being compared). Avg. shows the average within artificial or real faults. T.A. shows the total average across faults.} 
  \label{tab:main_results_fl_1_4_2}
   \scalebox{0.85}{
  \begin{tabular}{l|l|c|l|c|l}
    \toprule
    Id & GT  &\#F& Matches-GT  &\#M & DD-output \\
    \midrule    
M1 & WCI(0)  &1& 0  &0 & HLR  \\
M2 & ACH(7)  &1& 1  &\underline{1} & ACH(7)  \\
M3 & HLR  &1& 0  &0 & -  \\
C1 & ACH(2)  &1& 0  &0 & -  \\
C2 & HNE  &1& 0  &0 & -  \\
C3 & WCI(2)  &1& 0  &0 & -  \\
R1 & RAW(0)  &1& 0  &0 & -  \\
R2 & ACH(2)  &1& 1  &\underline{1} & ACH(2)  \\
R3 & HLR  &1& 0  &0 & -  \\
R4 & LCH  &1& 0  &0 & LRM | LAD | ACH(0)  \\
R5 & OCH  &1& 0  &0 & -  \\
R6 & WCI(0)  &1& 0  &0 & -  \\
R7 & ACH(2)  &1& 1  &\underline{1} & ACH(2)  \\
 \textbf{Avg.}& & \textbf{1}& & \textbf{0.2}&\\
    \midrule
D1 & ACH(7)  &1& 0  &0 & HLR  \\
D2 & OCH, HNE,  HBS  &3& 0, 0, 0  &0 & -  \\
D3 & OCH, LCH, ACH(0,1), HNE, HBS  &5& 0, 0, 0, 0, 0  &0 & -  \\
D4 & ACH(0,1), LCH, HLR  &3& 1, 0, 0  &1 & ACH(1)  \\
D5 & HNE, HBS  &2& 0, 0  &0 & -  \\
D6 & HLR, HNE, LCH, ACH(1)  &4& 0, 0, 0, 0  &0 & -  \\
D7 & HLR  &1& 0  &0 & -  \\
D8 & OCH, HLR  &2& 0, 0  &0 & -  \\
D9 & CPP, ACH(5,6), HBS  &3& 0, 0, 0  &0 & N/A  \\
 \textbf{Avg.}& & \textbf{2.7}& & \textbf{0.1}&\\
 \midrule
 \textbf{T.A.}& & \textbf{1.7}& & \textbf{0.2}&\\
\bottomrule
  \end{tabular}
  }
\end{table*}

\begin{table*}[htb]
  \centering
  \caption{Ground Truth (GT) and FL outcome generated by \textbf{\NL (NL)}; \#F indicates the number of ground truth faults, while \#M the number of ground truth faults detected by the tool (with underline used to indicate the best result among all tools being compared). Avg. shows the average within artificial or real faults. T.A. shows the total average across faults.} 
  \label{tab:main_results_fl_1_4_3}
   \scalebox{0.8}{
  \begin{tabular}{l|l|c|l|c|l}
    \toprule
    Id & GT  &\#F& Matches-GT  &\#M & NL-output \\
    \midrule    
M1 & WCI(0)  &1& 1  &1& WCI(0)  \\
M2 & ACH(7)  &1& 0  &0 & LCH  \\
M3 & HLR  &1& 0  &0 & N/A  \\
C1 & ACH(2)  &1& 0  &0 & -  \\
C2 & HNE  &1& 0  &0 & -  \\
C3 & WCI(2)  &1& 1  &\underline{1} & WCI(3)  \\
R1 & RAW(0)  &1& 0  &0 & -  \\
R2 & ACH(2)  &1& 0  &0 & LCH  \\
R3 & HLR  &1& 0  &0 & N/A  \\
R4 & LCH  &1& 1  &\underline{1} & LCH  \\
R5 & OCH  &1& 0  &0 & -  \\
R6 & WCI(0)  &1& 1  &\underline{1} & WCI(0)  \\
R7 & ACH(2)  &1& 0  &0 & LCH  \\
 \textbf{Avg.}& & \textbf{1}& & \textbf{0.3}&\\
    \midrule
D1 & ACH(7)  &1& 0  &0 & LCH  \\
D2 & OCH, HNE,  HBS  &3& 0, 0, 0  &0 & -  \\
D3 & OCH, LCH, ACH(0,1), HNE, HBS  &5& 0, 1, 1, 0, 0  &\underline{2} & ACH(1), LCH, LCN(0)  \\
D4 & ACH(0,1), LCH, HLR  &3& 1, 0, 0  &1 & ACH(0), BCI(0,1)  \\
D5 & HNE, HBS  &2& 0, 0  &0 & LCF(0)  \\
D6 & HLR, HNE, LCH, ACH(1)  &4& 0, 0, 0, 0  &0 & -  \\
D7 & HLR  &1& 0  &0 & N/A  \\
D8 & OCH, HLR  &2& 0, 0  &0 & -  \\
D9 & CPP, ACH(5,6), HBS  &3& 0, 0, 0  &0 & ACH(0), LCN(2,3)  \\
 \textbf{Avg.}& & \textbf{2.7}& & \textbf{0.3}&\\
 \midrule
 \textbf{T.A.}& & \textbf{1.7}& & \textbf{0.3}&\\
\bottomrule
  \end{tabular}
  }
\end{table*}

\begin{table*}[htb]
  \centering
  \caption{Ground Truth (GT) and FL outcome generated by \textbf{\UM (UM)}; \#F indicates the number of ground truth faults, while \#M the number of ground truth faults detected by the tool (with underline used to indicate the best result among all tools being compared). Avg. shows the average within artificial or real faults. T.A. shows the total average across faults.} 
  \label{tab:main_results_fl_1_4_4}
   \scalebox{0.85}{
  \begin{tabular}{l|l|c|l|c|l}
    \toprule
    Id & GT  &\#F& Matches-GT  &\#M& UM-output \\
    \midrule    
M1 & WCI(0)  &1& 0  &0 & HLR  \\
M2 & ACH(7)  &1& 1  &\underline{1}& ACH(7), HLR  \\
M3 & HLR  &1& 0  &0 & -  \\
C1 & ACH(2)  &1& 0  &0 & -  \\
C2 & HNE  &1& 0  &0 & -  \\
C3 & WCI(2)  &1& 0  &0 & -  \\
R1 & RAW(0)  &1& 0  &0 & -  \\
R2 & ACH(2)  &1& 1  &\underline{1} & ACH(2)  \\
R3 & HLR  &1& 1  &\underline{1} & HLR  \\
R4 & LCH  &1& 0  &0 & -  \\
R5 & OCH  &1& 0  &0 & -  \\
R6 & WCI(0)  &1& 0  &0 & -  \\
R7 & ACH(2)  &1& 1  &\underline{1} & ACH(2)  \\
 \textbf{Avg.}& & \textbf{1}& & \textbf{0.3}&\\
    \midrule
D1 & ACH(7)  &1& 0  &0 & -  \\
D2 & OCH, HNE,  HBS  &3& 0  &0 & ACH(7)  \\
D3 & OCH, LCH, ACH(0,1), HNE, HBS  &5& 0, 0, 0, 0, 0  &0 & -  \\
D4 & ACH(0,1), LCH, HLR  &3& 1, 0, 1  &\underline{2} & ACH(0,1), HLR  \\
D5 & HNE, HBS  &2& 0, 0  &0 & -  \\
D6 & HLR, HNE, LCH, ACH(1)  &4& 0, 0, 0, 0  &0 & -  \\
D7 & HLR  &1& 0  &0 & -  \\
D8 & OCH, HLR  &2& 0, 0  &0 & -  \\
D9 & CPP, ACH(5,6), HBS  &3& 0, 0, 0  &0 & ACH(0,2,4)  \\
 \textbf{Avg.}& & \textbf{2.7}& & \textbf{0.2}&\\
 \midrule
 \textbf{T.A.}& & \textbf{1.7}& & \textbf{0.3}&\\
\bottomrule
  \end{tabular}
  }
\end{table*}

\begin{table*}[htb]
  \centering
  \caption{Ground Truth (GT) and FL outcome generated by \textbf{GPT-4}; \#F indicates the number of ground truth faults, while \#M the number of ground truth faults detected by the tool (with underline used to indicate the best result among all tools being compared). Avg. shows the average within artificial or real faults. T.A. shows the total average across faults.} 
  \label{tab:main_results_fl_openai}
   \scalebox{0.85}{
  \begin{tabular}{l|l|c|l|c|l}
    \toprule
    Id & GT  &\#F& Matches-GT  &\#M& GPT-4-output \\
    \midrule    
M1 & WCI(0)  &1& 1&1& WCI(0), HLR, HNE, RCD(6)\\
M2 & ACH(7)  &1& 1  &\underline{1}& ARM(7), HLR, HNE\\
M3 & HLR  &1& 1&1& HLR, HNE, RCD(3,6), HBS\\
C1 & ACH(2)  &1& 1&1& ACH(2), HBS\\
C2 & HNE  &1& 1&1& HNE\\
C3 & WCI(2)  &1& 1&1& WCI(2)\\
R1 & RAW(0)  &1& 1&1& HNE, RAW(0), RCD(1)\\
R2 & ACH(2)  &1& 1  &\underline{1} & ACH(2), HNE, HBS, LAD\\
R3 & HLR  &1& 1  &\underline{1} & HLR, HNE, RCD(1), LAD, HBS\\
R4 & LCH  &1& 1&1& LCH, HNE, RCD(1), CPP, LAD\\
R5 & OCH  &1& 1&1& OCH, HNE |LAD, RCD (1), HBS\\
R6 & WCI(0)  &1& 1&1& WCI(0), HNE, RCD(1), LAD\\
R7 & ACH(2)  &1& 1  &\underline{1} & ARM (2), HNE\\
 \textbf{Avg.}& & \textbf{1}& & 1&\\
    \midrule
D1 & ACH(7)  &1& 1&1& ACH(7), HNE\\
D2 & OCH, HNE,  HBS  &3& 0, 1, 1&2& HNE, HBS, RCD(1,3,5), LCN(0,2,4), CPP\\
D3 & OCH, LCH, ACH(0,1), HNE, HBS  &5& 0, 1, 1, 1, 1&4& LCH, LAD, HNE, HBS, ACH(0)\\
D4 & ACH(0,1), LCH, HLR  &3& 1, 1, 1&\underline{3}& LCH, HLR, WCI(0,1), ACH(0,1),\\
& & & & & HBS, LAD, LCN(0)\\
D5 & HNE, HBS  &2& 1, 1&\underline{2}& HNE, HBS, LCF(0), LAD\\
D6 & HLR, HNE, LCH, ACH(1)  &4& 0, 1, 0, 0&1& HNE, LAD \\
D7 & HLR  &1& 0  &0 & LAD, HNE, HBS\\
D8 & OCH, HLR  &2& 0, 0  &0 & ACH(2), LCH, CPP, LAD, VRM\\
D9 & CPP, ACH(5,6), HBS  &3& 1, 0, 0&1& CPP, VRM\\
 \textbf{Avg.}& & \textbf{2.7}& & \textbf{1.6}&\\
 \midrule
 \textbf{T.A.}& & \textbf{1.7}& & \textbf{1.2}&\\
\bottomrule
  \end{tabular}
  }
\end{table*}